\documentclass[structabstract]{aa}  
\usepackage{graphicx}
\usepackage{lscape}
\usepackage{longtable}
\usepackage{natbib}
\bibpunct{(}{)}{;}{a}{}{,}
\usepackage{txfonts}
\begin{document}
  \title{CS, HC$_3$N and CH$_3$CCH multi-line analyses \\towards starburst galaxies}

  \subtitle{The evolution of cloud structures in the central regions of galaxies\thanks{This work is based on observations with the IRAM 30-m telescope. IRAM is supported by INSU/CNRS (France), MPG (Germany) and IGN (Spain). }}

  \author{R. Aladro
         \inst{1},
         J. Mart\'in-Pintado\inst{2}, S. Mart\'in\inst{3, 4}, R. Mauersberger\inst{5}, E.
Bayet\inst{6}}

  \institute{Instituto de Radioastronom\'ia Milim\'etrica (IRAM),
             Avda. Divina Pastora, 7, Local 20, E-18012 Granada, Spain\\
             \email{aladro@iram.es}
        \and
            Centro de Astrobiolog\'ia (CSIC-INTA), Ctra. de Torrej\'on Ajalvir km 4,
E-28850 Torrej\'on de Ardoz, Madrid, Spain.
   \and
        European Southern Observatory, Alonso de C\'ordova 3107, Vitacura, Casilla 19001, Santiago 19, Chile.
   \and
	Harvard-Smithsonian Center for Astrophysics, 60 Garden St. 02138, Cambridge, MA, USA.
   \and
        Joint ALMA Observatory, Av. Alonso de C\'ordova 3107, Vitacura, Santiago, Chile.
       \and
        Department of Physics and Astronomy, University College London, Gower Street,
London WC1E 6BT, UK.
            }

  \date{Received  /  Accepted }


 \abstract
  {}
  {We aim to study the properties of the dense molecular gas towards the inner few 100\,pc of four
nearby starburst galaxies dominated both by photo dissociation regions (M\,82) and large-scale
shocks (NGC\,253, IC\,342 and Maffei\,2), and to relate the chemical and physical properties of
the molecular clouds with the evolutionary stage of the nuclear starbursts.}
  {We have carried out multi-transitional observations and analyses of three dense gas molecular tracers,
CS, HC$_3$N (cyanoacetylene) and CH$_3$CCH (methyl acetylene), using Boltzmann diagrams in
order to determine the rotational temperatures and column densities of the dense gas, and using a Large Velocity Gradients model to
calculate the H$_2$ density structure in the molecular clouds.}
  {The CS and HC$_3$N data indicate the presence of density gradients in the molecular
clouds. These two molecules show similar excitation conditions, suggesting that they arise from the
same gas components. In M\,82, CH$_3$CCH has the highest fractional abundance determined in a extragalactic source ($1.1\times10^{-8}$). 
}
 {The density and the chemical gradients we have found in all galaxies can be explained
in the framework of the starburst evolution, which affects the chemistry and the structure of  molecular clouds around the galactic nuclei. The young shock-dominated
starburst galaxies, like presumably Maffei\,2, show a cloud structure with a rather uniform density
and chemical composition which suggests low star formation activity. Molecular clouds in
galaxies with starburst in an intermediate stage of evolution, such as NGC\,253 and IC\,342,
show clouds with a large density contrast (two orders of magnitude) between the denser regions (cores) and the less dense regions (halos) of the molecular clouds and relatively constant chemical abundance. Finally, the galaxy with the most evolved starburst, M\,82, has clouds with a rather uniform density structure, large envelopes of atomic/molecular gas subjected to UV photodissociating radiation from young star clusters, and very different chemical abundances of HC$_3$N and CH$_3$CCH.}

\authorrunning{Aladro et al. (2010)}
\titlerunning{CS, HC$_3$N and CH$_3$CCH towards the nuclei of starburst galaxies}
  \maketitle{}


\section{Introduction}
\label{intro}
Molecular clouds within the central region of our galaxy provide good templates for a  better understanding of the molecular emission from the nuclei of external galaxies. The study of well known molecular clouds in the Milky Way, like those affected by shock waves due to cloud-cloud collisions, mass loss from massive stars, gas accretion and explosive events like supernovae (e.g. \citealt{Jesus97}; \citealt{Huette98}), those  pervaded by UV dissociating radiation from the massive stars that create large photo dissociation regions (PDRs) (e.g. \citealt{Martin08a}), and also those which study regions affected by X-rays (XDRs) (e.g. \citealt{Jesus00}; \citealt{Arancha09}), allows to use the chemical composition to infer the physical processes dominating the heating of the molecular clouds. Such studies also provided a powerful tool to study not only the type of dominant activity but the evolution of the dense ISM in the obscured regions within galactic nuclei. For instance, the HCO$^+$-to-HCN line intensity ratios have been used for discriminating between AGNs and starburst signatures in the galaxy centers (\citealt{Kohno01,Imanishi04,Krips08}). Other intensity ratios, like HNC/HCN, are also used for differentiating between XDRs and PDRs contribution in different stages of evolution \citep{Loenel08, Baan08}. As another example which uses column densities derived from multi-line analyses, the HNCO/CS abundance ratio was proposed for differentiating between the starbursts mainly dominated by shocks or UV fields. (\citealt{Martin09}). 

The nearby ($\sim$3\,Mpc) galaxies NGC\,253 and M\,82, are two of the brightest infrared extragalactic sources, and among the most outstanding in terms of detection of extragalactic molecules (\citealt{MauerHenkel93,Martin06b}, Aladro et al., 2010b, in prep.). Simple molecules such as CO or CS are observed towards both sources with similar column densities and abundances. However, nearly all the species detected so far, in particular the more complex ones (NH$_3$, HNCO or CH$_3$OH), as well as SiO, show systematically lower abundances in M\,82 than in NGC\,253 and other starburst galaxies such as IC\,342 and Maffei\,2 \citep{Takano03,Martin09}. The observed differences among galaxies are not only due to excitation, but reflect different chemical compositions. The large overall abundance of HCO in M\,82 has been interpreted as evidence for a chemistry mainly dominated by photon-dominated regions (PDRs) within its central $\sim$\,650\,pc (\citealt{Burillo02}), which does not mean that other processes like shocks or X-ray heating do not take place in its nucleus \citep{Baan08}. The dissociating radiation explains the low abundance in M\,82 of the complex and fragile species cited above  (\citealt{Mauers03}; \citealt{Martin06}). Other molecules like CS are not under-abundant in M\,82 relative to NGC\,253, which can also be explained in terms of PDRs chemistry, since a high abundance of this molecule is predicted for slightly shielded regions (\citealt{Drdla89,Sternberg95}). This evidence, as well as the detection of other PDR enhanced species such as HOC$^+$ and CO$^+$ has led to think of M\,82 as the prototype of a giant extragalactic PDR (\citealt{Nguyen89,Mao00,Burillo02,Martin09}). 

\begin{table*}[!t]
\tiny
\caption[]{Main properties of the galaxies in our study.}
\begin{center}
\begin{tabular}[t!]{lccccccccccc} 
\multicolumn{7}{c}{}\\ 
\hline
Galaxy & RA & DEC & Type & D$\,^{\rm a}$ & $V_{\rm LSR}\,^{\rm b}$ & $\theta_{\rm s}\,^{\rm c}$ &  N(H$_2$)\,$^{\rm d}$ & $L_{\rm{IR}}$\, $^{\rm e}$& SFR$\,^{\rm f}$ & $M_{\rm{HI}}$\, $^{\rm e}$ & M$_{gas}^{\rm \, g}$\\
    & (J2000) & (J2000) &   & (Mpc) & (kms$^{-1}$)   &  ($''$)  &(x10$^{22}$\,cm$^{-2}$) &  (x10$^{9}$\,$L_\odot$)  & ($M_\odot$ yr$^{-1}$) & (x10$^{9}$\,$M_{\odot}$) & (x10$^8$\,$M_\odot$)\\
\hline
M\,82 &09:55:51.9 & +69:40:47.1 & I0 & 3.6 & 300 & 12.0 & 7.9 & 29.74 & $\sim$9 & 1.30 & 5.2\\
NGC\,253 & 00:47:33.4 & -25:17:23.0 & SAB(s)c & 3.9 & 250 & 20.0 & 6.2 &  15.10   & 3.6 & 2.57 & 28.0\\
IC\,342 & 03:46:48.5 & +68:05:46.0 & SAB(rs)cd & 3.3 & 31 & 10.6 & 5.8 & 2.26   & 2.5 & 18.2 & 1.4\\
Maffei\,2 & 02:41:55.1 & +59:36:15.0 & SAB(rs)bc & 3.3 & -17 & 11.0 & 4.4 & 2.7 & 0.5 & ... & 8.0\\
\hline
\end{tabular}
\begin{list}{}{}
\item[$^{\mathrm{a}}$] Values from \citet{Freedman94} for M\,82, \citet{Kara03} for NGC\,253 and IC\,342 and \citet{Fingerhut07} for Maffei\,2. 
\item[$^{\mathrm{b}}$]Velocities respect to the Local Standard of Rest. In M\,82 the value refers to the North-East lobe.
\item[$^{\mathrm{c}}$] Sizes of the emitting region, taken from: \citet{Martin06} for M\,82; This work for NGC\,253; \citet{Bayet06} for IC\,342; \citet{Mauers89b} for Maffei\,2. 
\item[$^{\mathrm{d}}$] Values taken from \citet{Mauers03}, and further corrected by the source size in column 7.   
\item[$^{\mathrm{e}}$] Values taken from \citet{Vaucou91}.
\item[$^{\mathrm{f}}$] Values taken from \citet{Strickland04} for NGC\,253 and M\,82; \citet{Walker92} for IC342.
\item[$^{\mathrm{g}}$] Estimation of the global molecular gas mass contained in the source. We have assumed that the gas is in a virial equilibrium. For the estimations we used the source sizes $\theta_{\rm s}$ of column 7 and an average linewidths calculated from those presented in Table~\ref{TableA1}.
\end{list}
\label{tab.Galaxies}
\end{center}

\end{table*}

On the other hand, the central regions of NGC\,253 and Maffei\,2 appear to resemble quite closely the giant molecular cloud complexes within our Galactic central regions where the low velocity shocks and massive star formation dominate the chemistry of the molecular material (\citealt{Martin06b}). Through the HNCO/CS ratio it has been shown that, the nuclear region of NGC253, though less dominated by UV radiation than in M\,82, is significantly more UV pervaded than in IC\,342 or Maffei\,2. This has been confirmed by the recent detection of the PDR tracers HCO, HOC$^+$, and CO$^+$ in NGC\,253 (\citealt{Martin09b}). The observed chemical differences are interpreted as the evolution of the starburst phenomenon, where M\,82 would be in a later stage (probably in a post-starburst phase where almost all the gas  has already been converted into stars), while NGC\,253 or Maffei\,2 centers would host younger starbursts (where still there are a large amount of reservoir gas which is being used to form the stars). This point can be used to link the stage of evolution of the starburst galaxy centers with the dominant physical processes (PDRs, shocks, XDRs), in such a way that a possible scenario, as seen by the single dish telescopes, could be that the more evolved (or post-starbursts) galaxies, like M82, are mainly dominated by UV fields creating large PDRs, while the younger starburst are basically dominated by shocks between the molecular clouds. It does not necessary implies that these are the unique processes taking place, but the dominant ones.

Interferometric maps of several molecules towards the center of IC\,342 show that both, PDRs and shock-dominated regions, can be resolved within its inner few 100\,pc (\citealt{Meier05}), 
while observations with single dish telescopes do not have the resolution to distinguish between these two types of activities in the nucleus. Thus, the overall scenario for IC342, based on beam averaged observations, is a mainly shock-dominated nucleus, as shown by the HNCO/CS ratio (\citealt{Martin09}). This galaxy seems to be in an early starburst stage, as shown by some shocks tracers like SiO or CH$_3$OH  (\citealt{Usero06}). On the other hand, Maffei\,2 seems the galaxy in our sample where C-shocks produced by cloud-cloud collisions play the most important role, being able to explain its high gas kinetic temperatures and chemical abundances (\citealt{Mauers03}). 

In this paper we present observations of CS, HC$_3$N and CH$_3$CCH towards the nuclei of four starburst galaxies. We aim to relate the chemical and physical properties derived from the analyses of these three dense gas tracers (densities, temperatures and fractional abundances) with the stage of evolution of each starburst. The outline of the paper is as follows. In Sect.~\ref{sect.Obs} we present our observations and other complementary line data taken from the literature. In Sect.~\ref{sect.Results} we explain the data analysis. The data are first analyzed under the local thermodynamic equilibrium (LTE) approximation in Sect.~\ref{sect.LTE}, where we obtain the column densities and rotational temperatures, which give us a first hint on the cloud structure. Sect.~\ref{sect.nonLTE} presents the results from a LVG modelling under non-LTE assumption. From this model, we estimate the H$_2$ densities for different cloud components. In Sect.~\ref{discussion}, we discuss the results of both approaches, taking into account their validity and limitations, and relating those results with the evolutionary stage of the molecular clouds in  each galaxy. Finally, in Sect.~\ref{conclusions} we present our conclusions.

\begin{table*}[!t]
\caption[]{Observed Lines.}
\begin{center}
\begin{tabular}[t!]{lrcccl} 
\multicolumn{5}{c}{} \\

\hline
Transition	& Frequency 	   & $F_{\rm eff}$/$B_{\rm eff}$	 & $\theta_{\rm beam}$ & $n_{\rm crit}$ $^a$ &Galaxy\\
 	    &	(GHz)	&		&	($''$)  	& (cm$^{-3}$)		\\
\hline
CS\,$(2-1)$ &	97.981	& 1.25 & 25.5	 & $1.2\times10^5$ &  NGC\,253, IC\,342 \\
CS\,$(3-2)$ &	146.969	& 1.35 & 16.8	 & $2.0\times10^6$ & IC\,342, M\,82 \\
CS$(5-4)$   &	244.936	& 1.79 & 10.1	 & $4.6\times10^7$ & NGC\,253, IC\,342, Maffei\,2, M\,82 \\
\hline
HC$_{3}$N\,$(9-8)$ &	81.881	& 1.22	& 30.2  & $1.6\times10^6$ & \emph{IC\,342}, \emph{Maffei\,2} \\
HC$_{3}$N\,$(10-9)$ &     90.979	& 1.23  & 27.5	 & $3.6\times10^6$ &   IC\,342 \\ 
HC$_{3}$N\,$(11-10)$ &	100.076	& 1.25	& 24.9 & $4.8\times10^6$ & \emph{IC\,342} \\ 
HC$_{3}$N\,$(12-11)$ &	109.174	& 1.27 & 22.2  & $7.2\times10^6$ & \emph{IC\,342}, \emph{Maffei\,2} \\
HC$_{3}$N\,$(15-14)$ &	136.464 & 1.32	& 18.2 & $2.6\times10^7$ & \emph{IC\,342},  \emph{M\,82}, \emph{Maffei\,2} \\
HC$_{3}$N\,$(16-15)$ &	145.561	& 1.35	& 16.9 & $7.0\times10^7$ &IC\,342,  \emph{M\,82} \\
HC$_{3}$N\,$(17-16)$ &	154.657	& 1.38	& 16.0 & $1.0\times10^8$ &\emph{IC\,342},  \emph{M\,82} \\
HC$_{3}$N\,$(18-17)$ &	163.753 & 1.41	& 15.1 & $1.3\times10^8$   &\emph{M\,82}\\
HC$_{3}$N\,$(23-22)$ &	209.230	&  1.60     & 12.0 & ... & Maffei\,2$^\star$ \\
HC$_{3}$N\,$(24-23)$ &	218.325	& 1.66  & 11.5 & ... &\emph{IC\,342}, \emph{M\,82}\\
HC$_{3}$N\,$(28-27)$ &	254.699	& 1.92	& 9.7 & ... & \emph{M\,82}, Maffei\,2$^\star$\\ 
\hline	
CH$_{3}$CCH\,$(5_0-4_0)$ &    85.457  &  1.22 & 29.1 & $3.7\times10^4$ & \emph{IC\,342}, \emph{Maffei\,2} \\
CH$_{3}$CCH\,$(6_0-5_0)$ &    102.548	&  1.28 & 24.2 & $1.3\times10^5$  &\emph{IC\,342}, \emph{Maffei\,2} \\
CH$_{3}$CCH\,$(8_0-7_0)$ &    136.728 & 1.32	&18.2 & $6.8\times10^5$  &\emph{M\,82}, \emph{IC\,342}, \emph{Maffei\,2} \\
CH$_{3}$CCH\,$(9_0-8_0)$ &    153.817 & 1.38	& 16.1 & $1.1\times10^6$  &\emph{M\,82} \\
CH$_{3}$CCH\,$(10_0-9_0)$ &   170.906 & 1.44	& 14.4 & $3.1\times10^6$  & \emph{M\,82} \\
CH$_{3}$CCH\,$(13_0-12_0)$ &  222.167 & 1.25\,/\,1.68\,$^b$     & 21.7\,/\,11.3\,$^b$ & $1.3\times10^7$  & NGC\,253, Maffei\,2$^\star$ \\
CH$_{3}$CCH\,$(14_0-13_0)$ &  239.252 &  1.80     & 10.3 &  $2.9\times10^7$  & Maffei\,2$^\star$\\
CH$_{3}$CCH\,$(15_0-14_0)$ &  256.337 & 1.87	& 9.6 & $4.4\times10^7$ & \emph{M\,82} \\
CH$_{3}$CCH\,$(16_0-15_0)$ &  273.420 & 2.02	& 9.1 & $9.4\times10^7$  &\emph{M\,82} \\
\hline
c-C$_3$H$_2$\,$(2_{1,2}-1_{0,1})$	&85.339	& 1.22	& 29.2	& ...	& \emph{Maffei\,2}	\\
c-C$_3$H$_2$\,$(5_{2,4}-4_{1,3})$ &  218.160	& 1.66	& 11.5	& ...	& \emph{M\,82}	\\
H$_2$CO\,$(2_{0,2}-1_{0,1})$	     &	145.603	& 1.35	& 16.9	& $2.3\times10^6$& M\,82, IC\,342	\\
H$_2$CO\,$(3_{0,3}-2_{0,2})$	     &	218.222	& 1.66	& 11.5	& $9.4\times10^6$	& M\,82, IC\,342	\\
H$_2$CO\,$(3_{2,2}-2_{2,1})$	     &	218.476	& 1.53	& 11.5	&  $5.2\times10^6$	& M\,82	\\
\hline
\end{tabular}
\begin{list}{}{}
\item[Last column indicates the galaxies where each transition was observed. Name in italics  means a new detection.]
\item[$^{\mathrm{*}}$]Name with $^\star$ means an upper limit of the line intensity. \item[$^{\mathrm{a}}$]Critical densities corresponding to a $T_{\rm{kin}}$\,=\,90\,K. Due to a lack of collisional coefficients in our LVG model for CH$_3$CCH, the $n_{\rm crit}$ for this molecule were calculated using the collisional coefficients of CH$_3$CN, which are very similar. 
\item[$^{\mathrm{b}}$] The first value corresponds to NGC\,253 observed with the JCMT telescope, and the second value corresponds to Maffei\,2 observed with IRAM 30-m telescope.
\end{list}{}{}
\label{tab.obs}
\end{center}
\end{table*}

\section{Observations and data reduction}
\label{sect.Obs}

We have detected a total of 37 transitions of CS, CH$_{3}$CCH and HC$_{3}$N towards the nuclear regions of the starburst galaxies M\,82, NGC\,253, IC\,342 and Maffei\,2, plus 5 additional non-detections. 26 out of the 37 transitions are newly detected. Table~\ref{tab.Galaxies} shows the main characteristics of these galaxies and Table~\ref{tab.obs} lists the observed lines. We have used the IRAM 30-m telescope (Pico Veleta, Spain), except for the CH$_3$CCH\,$(13-12)$ line, which has been observed with the JCMT telescope (Mauna Kea, Hawaii, USA)\footnote {The James Clerk Maxwell Telescope is operated by The Joint Astronomy Centre on behalf of the Science and Technology Facilities Council of the United Kingdom, the Netherlands Organization for Scientific Research, and the National Research Council of Canada.}. The observations were carried out in several periods between 2005 and 2009. At the IRAM 30-m telescope, we used the, now decommissioned, ABCD SIS receivers tuned simultaneously in single sideband mode in the 1, 2 and 3\,mm bands. Observations were carried out in wobbler switching mode with a symmetrical throw of 220$''$ in azimuth and a switching frequency of 0.5\,Hz. As back-end, we used the $256\times4$\,MHz filter banks. The beam sizes for each frequency are given in Table~\ref{tab.obs}. The pointing was checked every one or two hours towards several standard pointing calibrators, with an accuracy of $\sim3''$. The spectra were calibrated with a standard dual load system. The image sideband rejection at the observed frequencies ranged from 10 dB to 32 dB. The JCMT observations were carried out during the summer period in 2005. Observations were performed in beam switched mode with a frequency of
1\,Hz and a beam throw of $2'$ in azimuth. We used the A3 receiver in double side band to observe the CH$_3$CCH $J=13-12$ transition at 222.1\,GHz. As spectrometer we used the now decommissioned Digital Autocorrelation Spectrometer (DAS) in wide band mode which provided a bandwidth of 1800\,MHz at a 1.5\,MHz spectral resolution. The telescope beam size was 22$''$. The pointing was also checked every hour, with an accuracy of $\sim$3$''$.


The observed positions of the galaxies are shown in Table~\ref{tab.Galaxies}. M\,82 observations were pointed towards the North-East molecular clump, where the photodissociating radiation is claimed to be stronger, as seen in the HCO interferometric maps of \citet{Burillo02} (offsets ($+13'',+7.5''$) with respect to its coordinates in Table~\ref{tab.Galaxies}). In some cases, emission from the center of the galaxy was picked up by the beam, in particular for the lower frequency transitions (bigger beams) where the spectra show a bump at the left of the lines. NGC\,253 was observed towards the same position where the 2\,mm frequency survey was carried out by \citet{Martin06b}. Maffei\,2 and IC\,342 were observed towards their centers. 

\begin{figure}[!t]
\begin{center}
       \includegraphics[angle=0,width=0.5\textwidth]{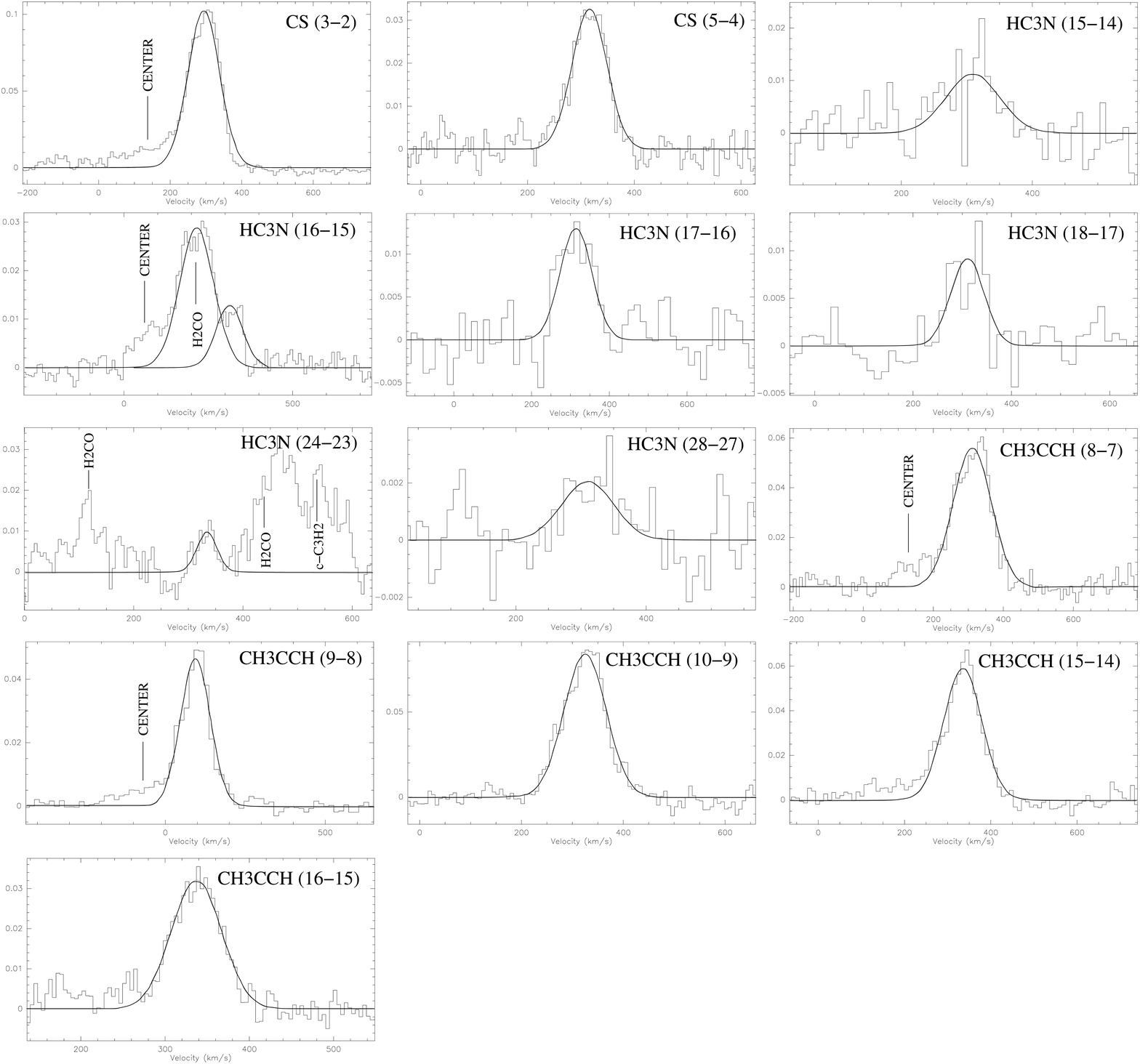}
       \caption{Gaussian profiles fits to observed lines in {\bf M\,82}. Temperatures are in $T_{\rm MB}$\,(K). Velocities are respect to the Local Standard of Rest. Some of the transitions show a bump located at the left of the line coming from the emission of the galaxy center. This feature is marked in the spectra with the word ``CENTER".}
  \label{MosGausM82}
\end{center}
\end{figure}

\begin{figure}[!t]
\begin{center}
	\includegraphics[angle=0,width=0.5\textwidth]{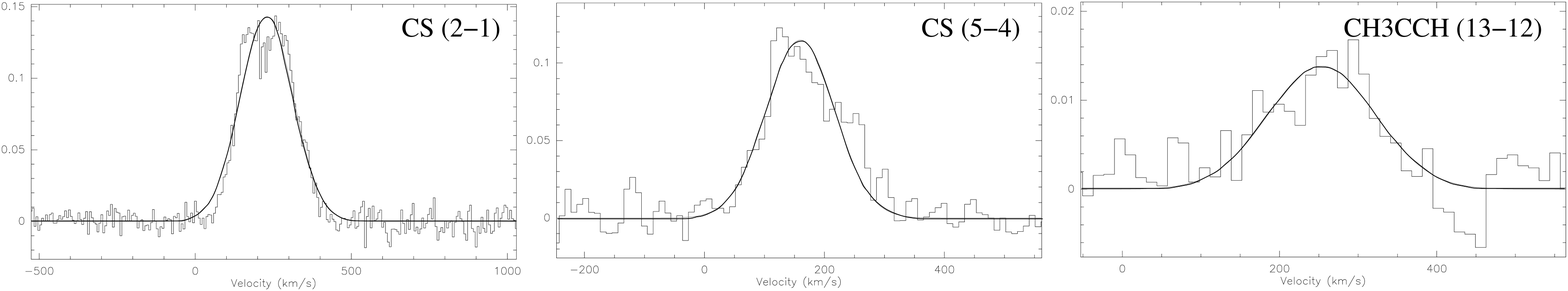}
       \caption{Gaussian profiles fits to observed lines in {\bf NGC\,253}. Temperatures are in $T_{\rm MB}$\,(K). Velocities are respect to the Local Standard of Rest.}
  \label{MosGausNGC253}
\end{center}
\end{figure}


\begin{figure}[!t]
\begin{center}

       \includegraphics[angle=0,width=0.5\textwidth]{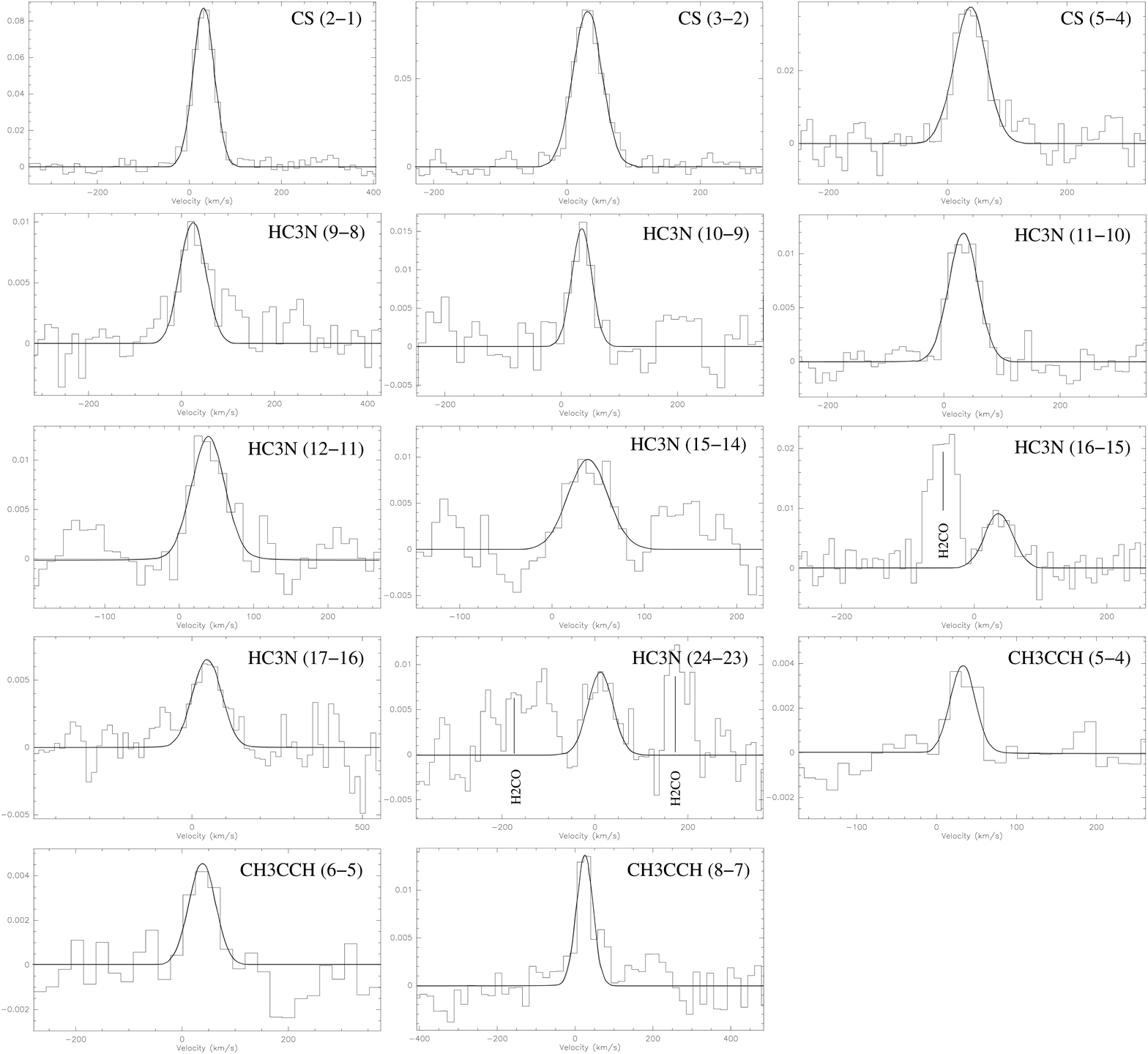}
       \caption{Gaussian profiles fits to observed lines in {\bf IC\,342}. Temperatures are in $T_{\rm MB}$\,(K). Velocities are respect to the Local Standard of Rest.}
  \label{MosGausIC342}
\end{center}
\end{figure}


\begin{figure}[!t]
\begin{center}
	\includegraphics[angle=0,width=0.5\textwidth]{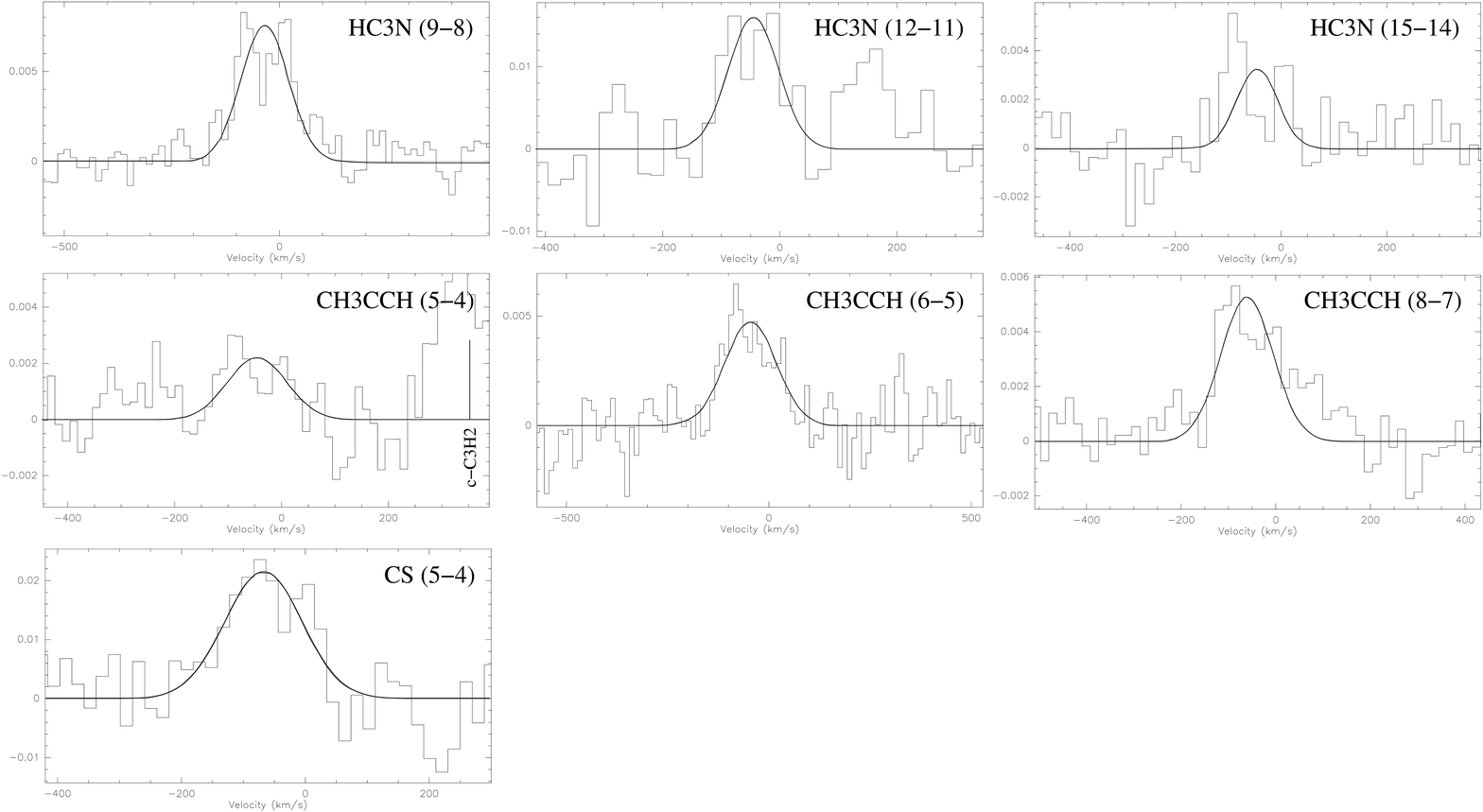}
       \caption{Gaussian profiles fits to observed lines in {\bf Maffei\,2}. Temperatures are in $T_{\rm MB}$\,(K). Velocities are respect to the Local Standard of Rest.}
  \label{MosGausMaffei2}
\end{center}
\end{figure}

The data, shown in Figs.~\ref{MosGausM82} to ~\ref{MosGausMaffei2}, are given in a $T_{\rm MB}$ scale, obtained from $T_A^*$ as $T_{\rm MB}$=($F_{\rm eff}$/$B_{\rm eff}$)\,$T_A^*$. The conversion factor ($F_{\rm eff}$/$B_{\rm eff}$) is given in Table~\ref{tab.obs} for each transition. Baselines of order 0 or 1 were subtracted in most cases. The rms of the residuals after the baseline subtraction are shown in Table~\ref{TableA1}. Gaussians profiles were fit to all detected lines (see Figs.~\ref{MosGausM82} to ~\ref{MosGausMaffei2}).
We have used the CLASS\footnote{CLASS $http://www.iram.fr/IRAMFR/GILDAS$} software package for data reduction and Gaussian profile fitting.

The intensity of the molecular emission for all species was also corrected by the beam dilution effect, due to the coupling between the source and the telescope beam, as $T_{\rm B}=[(\theta^2_{\rm s}+\theta^2_{\rm b})/\theta^2_{\rm s}]\,T_{\rm MB}$, where $T_{\rm B}$ is the source averaged brightness temperature, $\theta_{\rm s}$ is the source size, $\theta_{\rm b}$ is the beam size in arc seconds, and $T_{\rm MB}$ is the measured main beam temperature. Since CS, HC$_3$N and CH$_3$CCH require similar excitation conditions (see critical densities in Table~\ref{tab.obs}), it is plausible to use the same source size for all transitions. In the case of NGC253, we compared the brightness temperature of CS$(2-1)$ observed with the SEST telescope (\citealt{Martin05}) with the same line observed with the IRAM 30-m (this work). Using the relation between $T_{\rm MB}$ and $T_{\rm B}$, the comparison of both observations leads to a 20$''$ equivalent source size, which agrees with the value used by \citet{Martin06b}. For the other galaxies, there are no transitions observed at the same position with different telescopes, so we adopted source sizes derived from other studies listed in Table~\ref{tab.Galaxies}. The influence of the source size in the rotational temperatures and column densities can be seen in \citet{Bayet09} for CS in M\,82, NGC\,253 and IC\,342, among other galaxies. The assumption of similar source size for all transitions might have a small contribution to the derived temperature structure ($<30\%$) and less than a factor of 2 in the total column densities. On the other hand the assumed source size for each galaxy will have an almost insignificant effect on the temperatures ($<7\%$) but might have a strong effect in the column densities in the case of the source size being much smaller ($\theta_{\rm s}<2''$) than the one assumed.

Apart of CS, HC$_3$N and CH$_3$CCH, other molecules are detected in the observed bands. In particular, H$_2$CO\,$(2_{0,2}-1_{0,1})$, H$_2$CO\,$(3_{0,3}-2_{0,2}$ and H$_2$CO\,$(3_{2,2}-2_{2,1})$ have been detected in M\,82 and IC\,342.  c-C$_3$H$_2$\,$(5_{2,4}-4_{1,3})$ is seen in M\,82, and c-C$_3$H$_2$\,$(2_{1,2}-1_{1,0})$ is detected in Maffei\,2. These lines are shown in Figs.~\ref{MosGausM82} to ~\ref{MosGausMaffei2} and listed in Table~\ref{tab.obs}. No lines are blended except for HC$_3$N$(16-15)$, which is contaminated by  H$_2$CO\,$(2_{0,2}-1_{0,1})$. In this case, we estimated the contribution of  H$_2$CO\,$(2_{0,2}-1_{0,1})$ from other detections of this molecule (Aladro et al. 2010b, in prep.) and subtracted it. Then, we fitted a Gaussian profile to the residuals. 

\begin{figure*}
\begin{center}
       \includegraphics[angle=0,width=\textwidth]{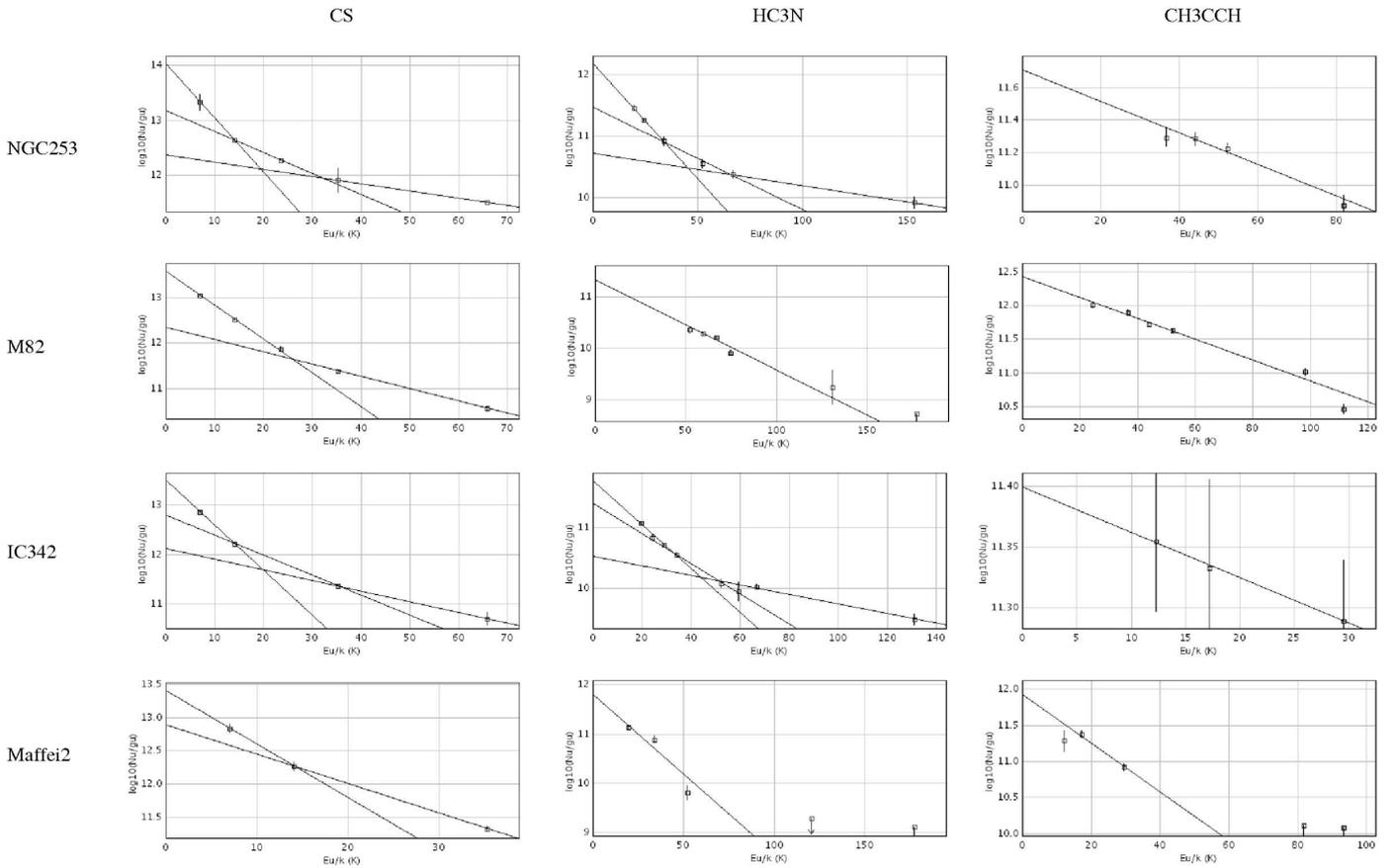}
       \caption{Boltzmann plots obtained with MASSA for all the galaxies and molecules. The column densities and rotational temperatures of each component were obtained from the linear fits. The errors correspond to those of the integrated intensities of the Gaussian profiles. The errors in the CS diagrams of NGC\,253 and Maffei\,2 also take into account the uncertainty due to the different observed positions used for each line, as detailed in Sect.~\ref{sect.LTE}.}
  \label{MosLTE}
\end{center}
\end{figure*}

In our analysis we also included some results from the literature in the 3, 2 and 1\,mm bands. We also included the CS$(7-6)$ sub-millimeter line when available. Details can be found in Table~\ref{TableA1}. For NGC\,253, we took the HC$_3$N lines from \citet{Mauersberger90}, that cover the full range from 81.9\,GHz to 236.5\,GHz. We also used several CH$_3$CCH lines from \citet{Martin06b} in the 2\,mm band. For M\,82, we complement our CS and CH$_3$CCH detections with those of \citet{Bayet09} and \citet{Mauersberger91}, respectively. Finally, for IC\,342 and Maffei\,2 only few CS lines were taken from the literature (see Table~\ref{TableA1}). It has to be noted that some CS lines compiled from the literature were observed towards slightly different positions, so there might be a small error associated to the different emission intensity in regions separated by a few arc seconds. In particular, the CS lines in NGC\,253 differ a maximum of 6$''$ in declination, while the CS lines in Maffei\,2 differ a maximum of 7$''$. In Sect.~\ref{sect.LTE},we estimate  the error introduced in our results by the difference in the observed positions.

We found that for NGC\,253 some of the compiled lines have a velocity resolution high enough for separating the two velocity components, roughly at 180 and 280 km\,s$^{-1}$, which arise from the molecular lobes separated by 10$''$, and located on both sides of the nucleus (\citealt{Harrison99}, \citealt{Mauers03}). For other observations, only one component was reported. In order to keep all data homogeneous, in our detections we have also fit only one Gaussian profile per line.  In any case, from the two velocity components analysis carried out by \citet{Bayet09} for CS and by \citet{Martin06b} for HC$_3$N, it can be seen that the differences between both components are of little significance.

In a similar way, Maffei\,2 also shows two velocity components that could not be separated in the lines taken from the literature. Therefore, in our detections we also fit only one Gaussian to both profiles. We have checked that a single Gaussian fit, both in NGC\,253 and Maffei\,2, leads to the same results as two Gaussians fit in terms of structure and properties of the gas. Thus, using just one velocity component does not affect our conclusions.


\begin{table*}
\caption[]{Parameters derived from the Boltzmann Diagrams.}
\begin{center}
\begin{tabular}[!h]{llccc} 
\hline
Galaxy	& Molecule 	   & N 	 & $T_{\rm{rot}}$ & N\,/\,N$_{\rm{H_2}}$\\
  &		&	(cm$^{-2}$) & (K)	\\
\hline
NGC \,253 &&& \\
\hline
 & CS$^\dagger$ & $(4.4\pm1.8)\times10^{14}$ &  4.4 $\pm$ 1.1 & $1.1\times10^{-8}$ \\
&     & $ (1.5\pm0.6)\times10^{14}$ & 12.0 $\pm$ 3 &  \\
&     & $ (0.7\pm0.3)\times10^{14}$ &  33.1 $\pm$ 8.3 &  \\
\cline{2-5}	         
& HC$_3$N &  $(8.1\pm2.9)\times10^{13}$ & 11.6 $\pm$ 1.8  & $2.2\times10^{-9}$ \\
&                &  $(3.6\pm1.4)\times10^{13}$  & 26.1 $\pm$ 4.5  &  \\
&                &  $(2.2\pm0.7)\times10^{13}$  & 73.3 $\pm$ 14.0  & \\
\cline{2-5}
& CH$_3$CCH  & $(3.2\pm1.0)\times10^{14}$ & 44.4 $\pm$ 7.7 & $5.2\times10^{-9}$ \\
\hline
M\,82 &&&\\
\hline
& CS & $(2.0\pm0.1)\times10^{14}$ & 5.8 $\pm$ 0.1 & $3.0\times10^{-9}$ \\
&          & $(3.6\pm0.5)\times10^{13}$ & 15.1 $\pm$ 0.9  &  \\
\cline{2-5}
& HC$_3$N & $(2.5\pm1.1)\times10^{13}$ & 24.7 $\pm$ 3.9 & $3.2\times10^{-10}$ \\
\cline{2-5}
 & CH$_3$CCH  & $( 8.5\pm0.9)\times10^{14}$ & 28.1 $\pm$ 1.2  & $1.1\times10^{-8}$ \\
\hline
IC\,342 && \\
\hline
& CS      & $(1.4\pm0.1)\times10^{14}$ & 4.7 $\pm$ 0.1 & $3.8\times10^{-9}$ \\
&         & $(6.0\pm0.1)\times10^{13}$ & 10.6 $\pm$ 0.2  &  \\
&         & $(2.3\pm0.1)\times10^{13}$ &  20.1 $\pm$ 4.0  &  \\

\cline{2-5}
& HC$_3$N     &  $(2.7\pm1.1)\times10^{13}$  & 13.1 $\pm$ 2.3  & $8.6\times10^{-10}$ \\
&             &  $(1.6\pm0.7)\times10^{13}$  & 20.6 $\pm$ 3.3  &  \\
&             &  $(7.2\pm3.2)\times10^{12}$  & 70.6 $\pm$ 21.2  & \\
\cline{2-5}
& CH$_3$CCH $^\star$ &  $4.5\times10^{14}$ & 70.0  & $7.8\times10^{-9}$  \\
\hline
Maffei\,2 &&\\
\hline
& CS$^\dagger$   & $ (1.2\pm1.0)\times10^{14}$ & 5.4 $\pm$ 4.5  & $4.3\times10^{-9}$ \\
&     & $(6.9\pm5.7)\times10^{13}$ & 9.7$\pm$8.1  &  \\
\cline{2-5}
& HC$_3$N  & $(4.3\pm0.6)\times10^{13}$ & 11.6 $\pm$ 0.8  & $9.8\times10^{-10}$ \\
&	   & $\le4.4\times10^{12}$       &     $\le$53.1      &               \\
\cline{2-5}
& CH$_3$CCH  & $(9.0\pm3.0)\times10^{13}$ & 12.8 $\pm$ 1.9 & $2.0\times10^{-9}$  \\
&	   & $\le1.1\times10^{14}$ &  $\le$46.9   & \\
\hline
\end{tabular}
\end{center}
\begin{list}{}{}
\item[Column densities (N) and rotational temperatures ($T_{\rm rot}$) obtained from the linear fits shown in Fig.~\ref{MosLTE} for each galaxy and molecule.] 
\item[The last column shows the relative abundances using the total $\rm H_{2}$ column densities of Table~\ref{tab.Galaxies}.] Before comparing, we have added up all the column densities we have obtained for a molecule.
\item[$^{\mathrm{\dagger}}$]The large errors in both N and $T_{\rm rot}$ reflect the uncertainty in the worse case scenario due to the different observed positions for the CS lines (see Sect.~\ref{sect.LTE}).  
\item[$^{\mathrm{\star}}$]The obtained values of $T_{\rm{rot}}$ and $N\rm{_{CH_3CCH}}$ are not reliable for the observed lines due to their large uncertainties (see the corresponding Boltzmann Diagram in Fig.~\ref{MosLTE}), so we have adopted $T_{\rm{rot}}$=70\,K and calculated $N_{\rm{CH_3CCH}}$ through this value.
\end{list}{}{}
\label{tab.Boltmann}
\end{table*}

\section{Results: Multi-transition analyses}
\label{sect.Results}

In this Sect. we present the detailed analysis of the multi-transition observations of CS, CH$_3$CCH, and HC$_3$N both under LTE and non-LTE approximations. During the analysis we derive the physical parameters for different molecular gas components within each source. It is important to note that these ``gas components'' are a simplification of the real structure of the molecular clouds. Though the real molecular clouds present gradients in temperature, density, and molecular abundances, this approach helps sketching the differences between the structure of molecular clouds in each galaxy.

\subsection{LTE analysis: Determination of rotational temperatures and column densities.}
\label{sect.LTE}
We analyzed the detected transitions under the LTE approximation, assuming optically thin emission. Estimations of the column density ($N$) and rotational temperature ($T_{\rm rot}$) were obtained using the Boltzmann diagrams (see \citealt{Goldsmith99} for a detailed explanation of the method and equations). The spectroscopic parameters were taken from the CDMS (\citealt{Muller01}; \citealt{Muller05}) and JPL catalogs (\citealt{Pickett98}).
 
In the case where all lines are emitted under LTE conditions with an uniform excitation temperature, a single straight line can be fitted to all population levels.
The derived $T_{\rm rot}$ will be a lower limit to the kinetic temperature ($T_{\rm kin}$) if the lines are sub-thermally excited, e.g. if the H$_2$ densities are not sufficiently high to counterbalance spontaneous decay of the excited levels, which is the assumption used here. However, molecular emission arises from clouds with  gradients in physical conditions which appears as different rotational temperatures components. In those cases, it is not possible to fit a single $T_{\rm rot}$ , but several values of $T_{\rm rot}$ are needed to fit all the data. Although a continuous temperature distribution seems more realistic, we here fit a discrete number of temperature components. This is sufficient to print out the temperature changes to be expected. In order to sample a wide range of physical conditions in the clouds, it is necessary to observe several transitions well separated in energies. This implies the observation of low and high excitation molecular lines. In this paper, we present the first detection of several high $J$ HC$_3$N and CH$_3$CCH transitions, which allow us to sample for the first time the high excitation gas components previously undetected. 

The Boltzmann diagrams for all galaxies and molecules are shown in Fig.~\ref{MosLTE}. The column densities and rotational tempera\-tures fitted through the LTE analysis are shown in Table~\ref{tab.Boltmann}. The column densities are source averaged with the source size of Table~\ref{tab.Galaxies}. Note that each N and $T_{\rm rot}$ are obtained from one linear fit, so they refer to two or more population levels. We used the MASSA\footnote{MASSA $http://damir.iem.csic.es/mediawiki-1.12.0/\\index.php/MASSA\_User's\_Manual$} software package  for this analysis. Relative abundances derived for the different gas components traced by every molecule are also shown in Table~\ref{tab.Boltmann}. 

As already pointed out in Sect.~\ref{sect.Obs}, the CS lines in NGC\,253 and Maffei\,2 were observed towards slightly different positions. We have estimated the error introduced in the column densities and rotational temperatures due to the differences in the observed positions. For NGC\,253, CS lines differ a maximum of 6$''$ in declination. In the worst case, this represents 60\% of the IRAM beam size. \citet{Mauers89} observed the CS$(2-1)$ line towards several positions of the NGC\,253 center, separated by $\pm$10$''$ both in declination and right ascension. Our ($\Delta$,\,$\Delta$)=$(0'',0'')$ position corresponds to their $(0'',+3'')$ position in their map. Then, we can calculate the intensity change due to an offset of 10$''$ in declination. Taking their $(0'',0'')$ and $(0'',+10'')$ positions, we obtain a decrease of 55\% in the integrated intensity and 34\% in the line intensity. Therefore, if the difference in declination is only of 6$''$, one could expect that the integrated intensity, used for plotting the Boltzmann diagrams, had to be decreased by less than 33\%. If we re-plot the Boltzmann diagram using $\pm$33\% of the integrated intensity of our CS$(2-1)$, the variation of the column density and rotational temperature is less than 41\% and 25\% respectively. These variations in $N_{\rm{CS}}$ and $T_{\rm rot}$ also reflect the normal deviation of the results between our work and that of \citet{Mauers89}, due to different instrumentation used and calibration errors.

Unfortunately, we cannot make such an estimate for the CS$(5-4)$ line, since it has not been observed towards several positions within the center of NGC\,253. However, for a separation of 6$''$ in declination, about half of the beam size at this frequency, we might expect a variation of less than a factor 1.3 in column densities and 1.2 in rotational temperatures. 

In the case of Maffei\,2, there are neither CS lines observed in several positions. The separation between the CS lines we have used is 7$''$ and 4$''$ in declination for the CS J=$2-1$ and $3-2$ respectively (1/4 of the beam in both cases, which means a variation of 15\% of the integrated intensity). Thus, the expected difference in column densities and rotational temperatures are of less than a factor 1.2.

The main results from the LTE analysis are:

\begin{itemize}
\item CS and HC$_3$N show similar excitation structure in NGC\,253 and IC\,342, where three temperature components can be fitted. This fact suggests that these two molecules are well mixed and tracing the same gas components. This possibility is further discussed and supported by the non-LTE analysis carried out in Sect.~\ref{sect.nonLTE}.

\item The excitation of the molecular gas in Maffei\,2 and M\,82 as traced by HC$_3$N can be fitted by a single rotational temperature. Only CS shows two different $T_{\rm rot}$.
The temperature gradient in these galaxies is significantly less pronounced than in NGC\,253 and IC\,342.
Nevertheless, since we cannot rule out higher excitation gas components, we computed upper limit values of $T_{\rm rot}$ and column densities from the undetected highest J-transitions of HC$_3$N and CH$_3$CCH in Maffei\,2 (see Table~\ref{TableA1}).

\item CH$_3$CCH shows a high excitation temperature relative to the other species for the galaxies in our sample ($T_ {\rm rot}$\,=\,13 and 44\,K for Maffei\,2 and NGC\,253 respectively, and 28\,K for M\,82). This is not conclusive for IC\,342, since the errors are still too large. These high rotational temperatures are expected from the relatively low dipole moment of CH$_3$CCH of 0.78 D (\citealt{Burrell80}), compared to 3.7 D for HC$_3$N (\citealt{DeLeon85}), which makes  $T_{\rm rot}$ from CH$_3$CCH a better approximation to  kinetic temperature than higher dipole molecules. Unlike other complex molecules such as CH$_3$OH, methyl acetylene shows high column densities and relative abundances in M\,82 ($\rm  N=8.5\times10^{14}$\,cm$^{-2}$, $ N/N_{\rm H_2}=1.1\times10^{-8}$). Since this molecule might be easily photo-dissociated by intense UV fields (\citealt{Fuente05}), lower abundances would have been expected in a PDR nucleus like the center of M\,82. 

\item The high $J$ lines of HC$_3$N trace the warmest gas, reaching rather high rotational temperatures of 73.3$\pm$14.0\,K in NGC253 and 70.6$\pm$21.2\,K in IC\,342. 
No other molecules have been found to have such high rotational temperatures in these galaxies so far, with the exception of those derived from ammonia of 120\,K and 443\,K for NGC\,253 and IC\,342, respectively. These results turn HC$_3$N into one of the best suited species for tracing the warmest and densest molecular gas in galaxies.

\end{itemize}

\subsection {Non-LTE analysis. Estimation of the H$_2$ densities.}
\label{sect.nonLTE}

In order to quantify the volume densities, n(H$_2$), and support/explain the multiple rotational temperatures, we have used a Large Velocity Gradient (LVG) statistical equilibrium model. Excitation effects due to line radiation trapping were taken into account, using the LVG approximation consisting of molecular clouds with homogeneous spherical distribution. In this model, we had two free parameters: the volume  density, $n(H_{\rm 2})$, and the column density per line width, $N$/$\Delta \rm v$. We assumed a background temperature to 2.7\,K, neglecting the contribution of any additional radiation field. Unfortunately, there are not collisional cross sections available for CH$_3$CCH so we could not run our LVG model in that case. Therefore, for this molecule we have used a LVG model of \citet{Mauersberger91}, which is explained in Sect.~\ref{LVG_results}. For CS and HC$_3$N, the collisional rates of our LVG model cover the first 12 energy levels for the first one and the first 23 energy levels for the second one. 

\begin{figure*}
\begin{center}
       \includegraphics[angle=0,width=0.85\textwidth]{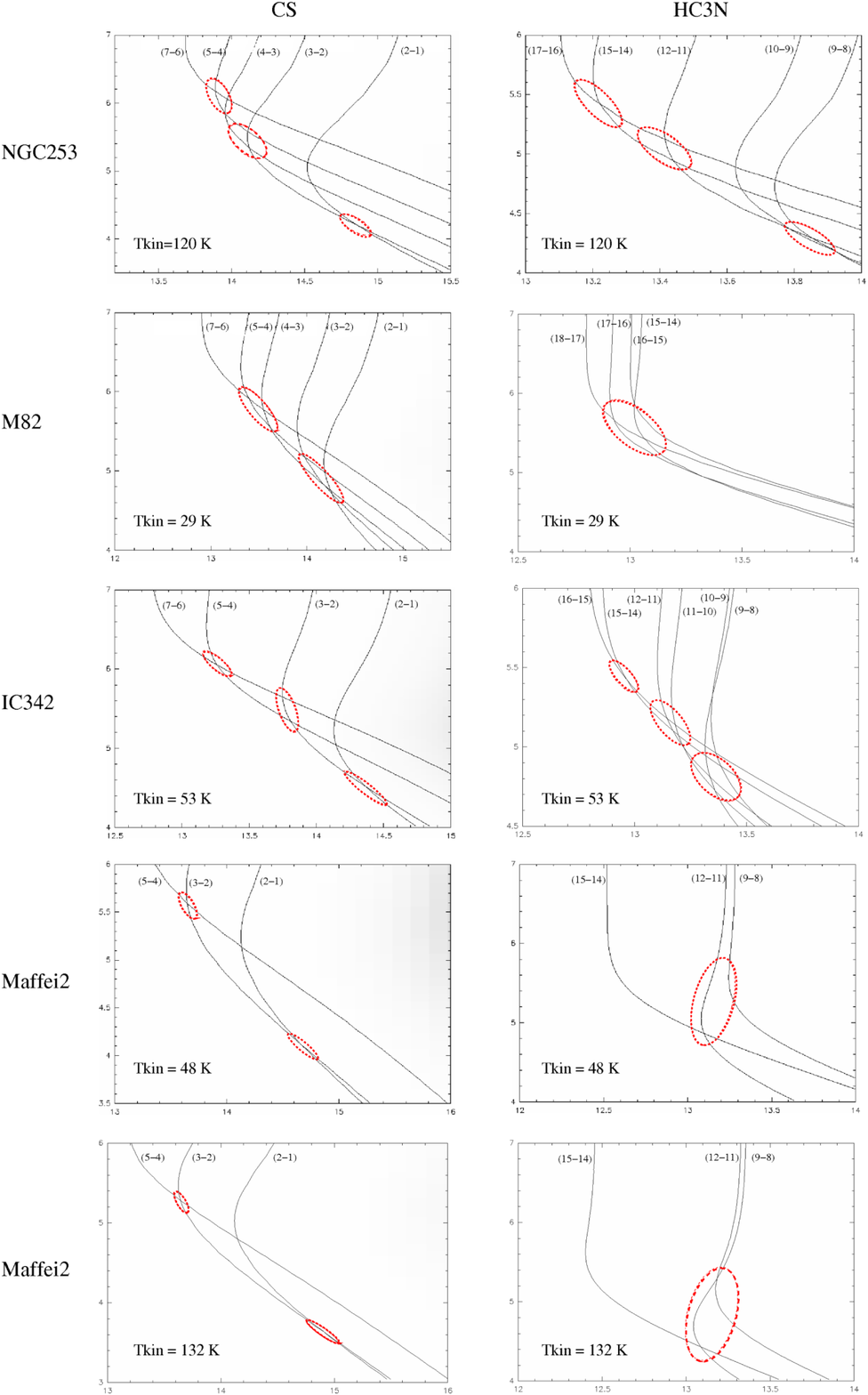}
       \caption{Results of the LVG model computations. The Y-axis is the log\,(n$_{\rm H_2}$) and the X-axis, the log\,(N), where N is the column density of the molecule. Each line represents the brightness temperature of a transition. When two or more lines cross, there is a density component. Elliptical shapes indicate where is more probable to find one of these density components within an uncertainty of  $T_{\rm b}\pm$10\% \,K.}
  \label{MosLVG}
\end{center}
\end{figure*}

\subsubsection{The kinetic temperature}
\label{Tkin}
LVG results suffer to some extent from a degeneracy between the derived $ n(H_{\rm 2}$) and the kinetic temperature. Some molecules, like ammonia (NH$_3$) or formaldehyde (H$_2$CO), are outstanding species usually used as thermometers, since the relative populations of their metastable levels are essentially only sensitive to the $T_{\rm kin}$. Considering that the emission from ammonia and the molecules in our study arise from similar regions, the LVG degeneracy can be broken, since we can use the additional information on the $T_{\rm kin}$  derived from ammonia observations by \citet{Mauers03}. The assumption of ammonia arising for a similar volume that the other molecules is justified by the fact that the large ammonia abundances found in these galaxies \citep{Mauers03} makes it very unlikely that a large fraction of the dense gas observed in CS and HC$_3$N does not contain NH$_3$.
The exception is M82, where the ammonia abundance is rather lower than in the clouds of the Milky Way. In this galaxy, H$_2$CO seems to arise from a similar volume than CO \citep{Mühle07}, which indicates that its molecular emission is more extended than that of NH$_3$, and also traces a warmer and more diffuse component in M\,82 than NH$_3$, CS, HC$_3$N or CH$_3$CCH. In fact, all these molecules have high dipole moments, need critical densities $ \sim10^4\,\rm cm^{-3}$ to be excited \citep{Mauers03} much larger than those derived from H$_2$CO \citep{Mühle07}. Thus, we find ammonia the most reliable tracer of the kinetic temperature, since its characteristics (dipole moment, critical density and emission size) are in an better agreement with CS, HC$_3$N or CH$_3$CCH than other tracers like CO and H$_2$CO.

\begin{table*}
\caption[]{Parameters obtained using our LVG code for CS and HC$_3$N, and the \citet{Mauersberger91} LVG model for CH$_3$CCH.}
\begin{center}
\begin{tabular}[!h]{llcccccccc} 
\hline
Galaxy	& $T_{\rm kin}$ & Molecule   & 	n(H$_2$)   & N                    & $N_{\rm{LTE}}/N_{\rm non-LTE}$  \\
        &       &     & (cm$^{-3}$)  & (cm$^{-2}$) &           \\
\hline
NGC\,253 & 120\,K &&&& \\
\hline
 &  &CS        & $(1.7\pm0.6)\times10^{4}$           &   $(7.3\pm1.7)\times10^{14}$   &    0.6         \\
  & &          & $(3.1\pm1.4)\times10^{5}$           &   $(1.3\pm0.3)\times10^{14}$   &    1.1         \\
  &&           & $(1.4\pm0.7)\times10^{6}$           &   $(8.0\pm2.0)\times10^{13}$   &     0.9       \\
\cline{3-6}
&  &HC$_3$N    & $(2.0\pm0.8)\times10^{4}$           &   $(7.2\pm1.1)\times10^{13}$   &  1.1            \\
&  &           & $(1.2\pm0.4)\times10^{5}$           &   $(2.6\pm0.5)\times10^{13}$   &   1.4          \\
&  &           & $(2.9\pm1.3)\times10^{5}$           &   $(1.7\pm0.3)\times10^{13}$   &     1.3       \\
\cline{3-6}
&  &	CH$_3$CCH	&$ \sim10^5$	&...	&...\\
&  &		&$\sim10^6$	&...	&...\\

\hline
M\,82 & 30\,K &&&& \\
\hline
& & CS         & $(1.1\pm0.4)\times10^{5}$           &   $(3.2\pm1.4)\times10^{14}$   &    0.6         \\
&  &           & $(2.3\pm1.6)\times10^{6}$           &   $(3.6\pm1.3)\times10^{13}$   &     1.0        \\
\cline{3-6}
&  &HC$_3$N    & $(1.7\pm1.3)\times10^{6}$           &   $(1.9\pm0.6)\times10^{13}$   &   1.4          \\
\cline{3-6}
&  &	CH$_3$CCH	&$\sim10^5$	&...	&...\\
\hline
IC\,342 & 53\,K &&&& \\
\hline
 &  &CS        & $(3.3\pm1.4)\times10^{4}$           &   $(2.5\pm0.9)\times10^{14}$   &    0.6        \\
 &  &          & $(3.4\pm1.8)\times10^{5}$           &   $(6.2\pm1.2)\times10^{13}$   &    1.0          \\
 &  &          & $(1.2\pm0.4)\times10^{6}$           &   $(1.9\pm0.5)\times10^{13}$   &    1.2           \\
\cline{3-6}
&  &HC$_3$N    & $(6.7\pm2.0)\times10^{4}$           &   $(2.4\pm0.6)\times10^{13}$   &     1.1     \\
&  &           & $(1.5\pm0.5)\times10^{5}$           &   $(1.5\pm0.3)\times10^{13}$   &     1.1       \\
&  &           & $(2.8\pm0.6)\times10^{5}$           &   $(9.1\pm1.3)\times10^{12}$   &     0.8     \\
\cline{3-6}
&  &	CH$_3$CCH	&$\sim10^5$	&...	&...\\
&  &		&$\sim10^6$	&...	&...\\
\hline                                                                
Maffei\,2 & 48\,K &&&& \\                                             
\hline
 &  & CS &$(1.3\pm0.4)\times10^{4}$ & $(5.2\pm1.5)\times10^{14}$ & 0.2 \\
 &  &   & $(3.8\pm1.1)\times10^{5}$ & $(4.6\pm0.9)\times10^{13}$ & 1.5 \\
\cline{3-6}                                                                
 &  &HC$_3$N        & $(3.4\pm2.9)\times10^{5}$           &   $(1.5\pm0.4)\times10^{13}$   &    2.9           \\
\cline{3-6}
&  &	CH$_3$CCH	&$\sim10^4$	&...	&...\\
\hline                                             
Maffei\,2 & 132\,K &&&& \\
\hline
 &  & CS & $(4.5\pm1.4)\times10^{4}$ & $(8.7\pm2.8)\times10^{14}$ & 0.1 \\ 
&  &     & $(1.8\pm0.6)\times10^{5}$ & $(4.6\pm0.9)\times10^{13}$ & 1.5 \\ 
\cline{3-6}
 &  &HC$_3$N        & $(1.3\pm1.1)\times10^{5}$           &   $(1.5\pm0.5)\times10^{13}$   &     2.9        \\
\cline{3-6}
&  &	CH$_3$CCH	& ...	& ...	& ...\\
\hline
\end{tabular}
\end{center}
\begin{list}{}{}
\item[The last column reflects the ratio between the column densities obtained by both LTE and non-LTE approaches.]
\end{list}{}{}
\label{tab.LVG}
\end{table*}

Thus, we have used the kinetic temperatures derived by \citet{Mauers03} as an input parameter in our LVG model. They obtained a $T_{\rm rot}$ in NGC\,253 of 142\,K and 100\,K, for the two velocity components. Since we do not make any distinction between these two velocity components, we have used the average value of  $T_{\rm kin}$=120\,K. In the case of M\,82, the kinetic temperature traced by ammonia was 29\,K. On the other hand, Maffei\,2 and IC\,342 show two different rotational temperatures, one for the lower $(J,K)$ transitions of ammonia, and another one for the higher $(J,K)$ transitions. In the case of IC\,342, the $T_{\rm rot}$ are of 53\,K  and 443\,K respectively. The first one is derived from the transitions $(J,K)$\,=\,$(1,1)$ to $(3,3)$, having energies in their Boltzmann diagram ranging from 0 to 150\,K, while the other component of 443\,K is obtained from the fit to the highest transitions, $(J,K)$\,=\,$(5,5)$ to $(9,9)$, with energies from 300\,K to more than 800\,K . In our IC342 population diagrams, the energy levels reach, as maximum, 130\,K. Thus, one could guess that the gas component at 443\,K has little contribution to the relatively low energy transitions we deal with in this paper. So, we have used $T_{\rm kin}$= 53\,K in our LVG analysis shown in Table~\ref{tab.LVG}.

For Maffei\,2, the two $T_{\rm kin}$ of 48\,K and 132\,K were taken into account when running the LVG model. Table~\ref{tab.LVG} shows the LVG results for both temperatures. Gas densities obtained from CS for this source are more sensitive to the changes of the $T_{\rm kin}$ than those obtained from HC$_3$N, with variations of up to one order of magnitude in the lower density component. Moreover, the CS and HC$_3$N transitions seem to trace gas with low excitation temperature, as shown by the LTE analysis in Sec~\ref{sect.LTE}  (from 5 to 10\,K for CS and 12\,K for HC$_3$N). Following the same arguments used for IC\,342 when comparing our upper energy values of the Boltzmann plots with those of \citet{Mauers03}, and taking into account that the rotational temperatures are considered as a lower limit of the kinetic temperature, it seems more appropriate to use $T_{\rm kin}$=48\,K instead of T$_{\rm kin}$=132\,K for Maffei\,2.

\subsubsection{LVG model results}
\label{LVG_results}
Fig.~\ref{MosLVG} shows the predictions from our LVG modeling for all galaxies in a $log[n({\rm H_ 2})]-log [N]$ diagram for CS or HC$_3$N and taking the kinetic temperatures given above. Lines represent the model results matching the observed brightness temperatures for each transition. The points where two or more lines intersect correspond  to the H$_2$ density and the column density which fit those lines. The dashed ellipses show the area where two or more lines intersect to within an error of $\pm$10\%.  Such conservative error was chosen taking into account the $T_{\rm kin}$ errors obtained by \citet{Mauers03}, which are always lower than 10\% (with the only exception of the 30\% uncertainty in the $T_{\rm rot}$\,=\,48\,K for Maffei\,2. Nevertheless, such error does not significantly affect our derived parameters). As shown in Fig. 6, we find that several components with different densities are needed to fit all observed lines (as already shown by \citealt{Bayet08} and \citealt{Bayet09}), in agreement with the multiple rotational temperatures obtained from the Boltzmann diagrams. Table~\ref{tab.LVG} shows the derived $n(\rm H_2)$ and $N$ for the different density components in all the galaxies.  The errors in H$_2$ densities and column densities are derived from the 1$\sigma$ errors in the Gaussian fits. The ratios between the column densities of CS and HC$_3$N derived from the LTE analysis and the LVG model for the different density components are also shown in the last column of Table~\ref{tab.LVG}.

Unfortunately, our LVG model does not include the collisional cross sections of CH$_3$CCH, and we do not have at our disposal any other similar code prepared to use it. So far, the best approximation to model the excitation of CH$_3$CCH was done by \citet{Mauersberger91}. They run a LVG model for methyl acetylene using the CH$_3$CN collisional rates and rotational constants, since these parameters may be similar for both species. Therefore, for methyl acetylene we used the results of \citet{Mauersberger91} as a {\emph {rough estimation}} of the $\rm H_2$ volume densities within an order of magnitude uncertainty, i.e. $\sim10^4$,$\sim10^5$ or $\sim10^6\,\rm cm^{-3}$. On the other hand, although the temperatures used in the \citet{Mauersberger91} model for CH$_3$CCH do not exactly match those assumed from NH$_3$, these differences in the $T_{\rm kin}$ do not play a relevant role and the volume densities  hardly vary when using either the ammonia or the methyl acetylene kinetic temperatures.
 
The main results from both LVG analyses can be summarized as follows:

\begin{itemize}
\item LVG analyses have been used to quantify the density and the molecular column densities of the gas components inferred from the Boltzmann plots. The column densities obtained by both approaches are consistent within a factor of 2 in almost all cases (see Table~\ref{tab.LVG}).

\item CS traces a very wide range of densities. Similar to the LTE results our data requires three density components in NGC\,253 and IC\,342, with a density contrast of up to two orders of magnitude between the more diffuse gas ($\sim10^4\rm cm^{-3}$) and the densest material ($\sim10^{6}\rm cm^{-3}$).
On the other hand, only two density components are needed to fit the data in Maffei\,2 and M\,82, with a smoother density gradient of about one order of magnitude.

\item  HC$_3$N traces a narrower range of densities than CS in M\,82 and Maffei\,2. In these sources, only one density component is found, with a density of a few 10$^6$ cm$^{-3}$ in M\,82, and a several 10$^5$ cm$^{-3}$ in Maffei\,2. In fact, cyanoacetylene is not found in the more diffuse gas ($\sim$10$^4$ cm$^{-3}$) observed in CS in these galaxies. For the other two galaxies, IC\,342 and NGC\,253, both species trace the same density components, but HC$_3$N  systematically provides lower  densities than those of CS.

\item Using the grid of models of \citet{Mauers03}, we have found that CH$_3$CCH seems to be arising from an intermediate component of $\sim10^5\,\rm cm^{-3}$ in M\,82, while for IC\,342 and NGC\,253 could be tracing gas of densities $ \sim10^5\,\rm cm^{-3}$ or/and $ \sim10^6\,\rm cm^{-3}$. In the case of Maffei\,2, methyl acetylene arises from less dense molecular gas of densities $ \sim10^4\,\rm cm^{-3}$.

\end{itemize}

\begin{figure*}
\begin{center}
	\includegraphics[angle=0,width=\textwidth]{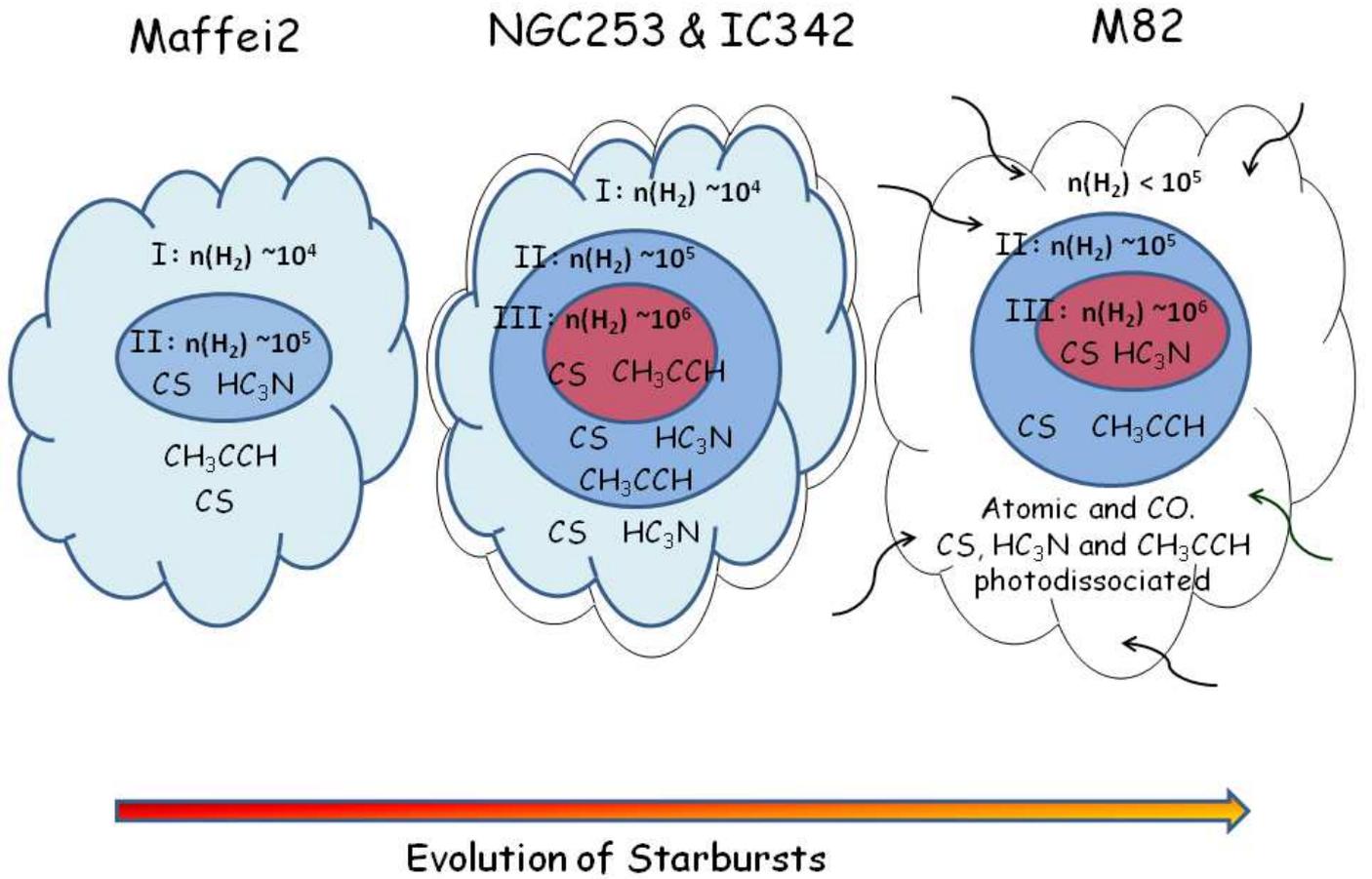}
	\caption{This simple cartoon shows how CS, HC$_3$N and CH$_3$CCH could be distributed in the giant molecular clouds within the central hundred parsecs of the four galaxies from the LVG model results. The size scales range from $\sim$145 to $\sim$560\,pc depending on the galaxy distance and beam size. We have found a Region I of densities about $\sim$10$^4 \rm cm^{-3}$, an intermediate Region II of densities $\sim$10$^5 \rm cm^{-3}$, and a Region III of high densities around $\sim$10$^6 \rm cm^{-3}$. Maffei\,2 shows two density components, but no core (Region III) is detected. In NGC\,253 and IC\,342, up to three gas components of different densities are distinguished ($\sim$10$^4$, $\sim$10$^5$ and $\sim$10$^6$\,cm$^{-3}$). M\,82 shows a region of intermediate densities and a core. The Region I is not detected by the molecules used here, although it should be surrounding the Region II.}
  \label{nubes1}         
\end{center}
\end{figure*}

\begin{table*}
\caption[]{N(HC$_3$N)\,/\,N(CS) ratios and percentage contribution of each Region to the overall dense molecular gas}
\begin{center}
\begin{tabular}[!h]{cccccccccc} 
\hline
                     &                   Maffei\,2 & IC\,342   & NGC\,253        & M\,82                  \\
                     &  \multicolumn{4}{c}{($N_{\rm HC_3N}/N_{CS}-\% \rm contribution$)}  \\
\hline
     Region I  ($\sim$10$^4$\,cm$^{-3}$)  &$...-92\% $ & $0.10-75\%$  & $0.10-77\%$   & $...-...   $   \\
     Region II ($\sim$10$^5$\,cm$^{-3}$)  & $0.33-8\% $ & $0.39-19\%$  & $0.33-14\%$   & $...-90\%   $     \\
     Region III ($\sim$10$^6$\,cm$^{-3}$) &$...-...  $ &$...-6\%  $  & $...-9\%  $   & $0.53-10\%   $      \\
\hline
\end{tabular}
\end{center}
\begin{list}{}{}
\item[The fist value indicates the N(HC$_3$N)\,/\,N(CS) ratio obtained from our LVG code.] The second value represents the percentage contribution of each Region (I, II or III)  to the overall dense molecular gas, traced by CS, within the molecular clouds in the center of the galaxies, obtained from our LVG model.
\end{list}{}{}
\label{tab.regions}
\end{table*}

\section{Discussion: Molecular cloud structure in the nuclei of galaxies}
\label{discussion}

In order to understand the physical structure and properties of the ISM of the galaxies in this paper, we will use the simple molecular cloud structures sketched in Fig.~\ref{nubes1}. It illustrates the expected evolution of molecular clouds in starburst galaxies. We consider three different regions: The diffuse outer region (halo) with densities of $\sim10^{4}\rm\, cm^{-3}$ referred to as {\emph {Region I}; the dense gas with densities of $\sim10^{5}\rm\, cm^{-3}$  referred to as {\emph {Region II}}; and the inner and densest gas we can trace (core), the {\emph {Region III}}, with densities of $\sim10^{6}\rm\,cm^{-3}$.} Fig.~\ref{nubes1} illustrates how the CS, HC$_3$N and CH$_3$CCH seem to be distributed in the molecular clouds of the central hundred parsecs of the galaxies in our sample. At first glance, the molecular clouds of NGC\,253 and IC\,342 have a very similar density distributions, while Maffei\,2 and M\,82 show a different structures. Next, we will discuss why these density distributions point to three different stages of the starburst evolution.

We can make a rough estimation of the molecular cloud sizes, which allows us to know how many molecular clouds could be contained in our beam. Supposing that the gas is in a virial equilibrium and formed by homogeneous clouds of uniform H$_2$ density of n\,=\,10$^5$\,cm$^{-3}$, we have obtained an approximate molecular clouds radii of $\sim$29, $\sim$51, $\sim$19 and $\sim$34\,pc for M\,82, NGC\,253, IC\,342 and Maffei\,2 respectively. Taking into account the source size used for each galaxy (see Table~\ref{tab.Galaxies}), it is possible to estimate the number of molecular clouds contained in the beam of about $\sim$7, $\sim$7, $\sim$9 and $\sim$5 for M\,82, NGC\,253, IC\,342 and Maffei\,2. Although there are not big differences in the numbers, they seem to be consistent with the proposed overall scenario: M82, the galaxy most dominated by the UV fields, is claimed to have smaller and more fragmented clouds \citep{Fuente08}. Maffei2, on the other hand is the galaxy most dominated by shocks and its clouds seem to be rather bigger, having less number of them in the beam ($\sim$5). However, these are just rough calculations, since we show in this paper the non negligible temperature and density gradients within molecular clouds, as depicted in Fig.~\ref{nubes1}. In fact, the observed gas could well be formed of several molecular clouds of different sizes each. Thus, this is something that could be better constrained using theoretical models or interferometric data. Also, the assumed distances to the galaxies play an important role in the values obtained. Nevertheless, in order to see whether our values are consistent with those of the literature, we have compared the most straightforward case of IC342. The size of the molecular clouds we have obtained in this galaxy and the HC$_3$N$(10-9)$ interferometric map of \citet{Meier05} are in good agreement, since they point out that the emission of HC$_3$N in IC\,342 is not expected to be more extended than $\sim$50\,pc.

On the other hand, from the analysis presented in Sect.~\ref{sect.Results} the CS abundances remain almost constant independently of the dominant process, with the exception of NGC\,253. This is likely due to the fact that CS is not significantly affected by any of the dominant process (PDRs or shocks), as shown by \citet{Requena06} in the Galactic Center, and \citet{Martin09} in a sample of galaxies.  On the other hand, HC$_3$N shows low abundance in M\,82. This molecule seems to be easily dissociated in PDRs like in the center of this galaxy, as was already proposed by \citet{Arturo98}. At the same time, HC$_3$N is absent in the non-detected Region I of M\,82, likely due to the lack of shielding from the dissociating radiation. This is consistent with the idea that CS is more resistant to the UV than HC$_3$N; and cyanoacetylene was not detected in the Region I in Maffei\,2 either. In this case, the most likely explanation is that the low density in this region does not excite the molecule, so  its lack could be due to excitation effects rather than to chemistry.

As shown in Table~\ref{tab.regions}, the HC$_3$N/CS ratio hardly varies when comparing regions with similar densities among the galaxies. 
This supports our sketch of the molecular clouds divided in regions, as they not only share similar densities but also molecular abundances. In Table~\ref{tab.regions} we also present the fraction of gas that each of these components contribute to the overall molecular gas traced by CS (we take this molecule as a reference since it is the one which traces a wider range of densities). In the following we discuss the structure of the different observed sources and its possible relation to the evolution of their nuclear starburst.

\subsection{Molecular clouds in galactic centers hosting young starburst. The case of Maffei\,2}

Maffei\,2 is considered to be at a very early state of the starburst evolution (\citealt{Martin09}), with a  large amount of reservoir gas in its center, and only a moderate star formation rate (Table~\ref{tab.Galaxies}). The gas kinetic temperatures derived from ammonia ($T_{\rm kin}$\,=\,48 and 132\,K from \citealt{Mauers03}) indicate that shocks dominate the ISM heating (as already indicated by e.g. \citealt{Ishiguro89}), which is consistent with the HNCO/CS ratio. There are several processes that can contribute to generate shocks in the early phases of the starbursts. One is the existence of a bar, which creates a potential leading to cloud-cloud collisions at the edges. This is, in fact, the case of Maffei\,2. The molecular cloud collisions can be also generated by the usual compression which leads to the star formation; and there could also be a contribution from the stellar winds of the newly formed stars. 

The molecular clouds in Maffei\,2 seem to have large halos (i.e., Region I, which contains $>90\%$ of the detected CS) of molecular gas very rich of fragile molecules, like for example ammonia or HNCO, but also CO, as seen by \citet{Israel03}. The molecular complexes of Maffei\,2 show low rotational temperatures of 5-15 K in the molecules studied, and normal fractional abundances of $\sim10^{-9}$ for CS, HC$_3$N and CH$_3$CCH (see Table~\ref{tab.Boltmann}), which are similar to those of the ``quiescent'' molecular clouds in the Galactic Center (GC), with no, or low, star formation activity. On the other hand, if we consider the upper limits to the undetected lines in Maffei\,2 in the LTE analysis, the derived rotational temperatures are less than 53\,K from HC$_3$N and 47\,K from CH$_3$CCH (see Table~\ref{tab.Boltmann}). 
 
Taking into account all the results from the LTE and non-LTE analyses, the three studied molecules seem to be arising from molecular clouds similar to the molecular clouds with low star formation activity found in the central region of the Milky Way.
At this stage, the molecular material is likely to be funneled towards the center of the potential well. In fact, a bar of molecular gas drives the material to the inner central parts \citep{Ishiguro89,Hurt91,Kuno08}, increasing its density and feeding the still relatively low star formation rate found in Maffei\,2.

\subsection{ Molecular clouds in galaxies hosting intermediate-age starbursts: NGC\,253 and IC\,342} 

The two galaxies in our sample, NGC\,253 and IC\,342, which are considered to host a starburst at intermediate stage of evolution, where a large amount of gas has been already converted into starts (\citealt{Rieke88}; \citealt{Schinnerer2008}; \citealt{Martin09b}), show molecular clouds with similar physical and chemical structure (see Table~\ref{tab.regions}). The median age of the star clusters in the center of NGC253, studied at optical wavelengths, is $\sim$6\,Myr \citep{Onti09}, while from near infrared (NIR) continuum emission, \citet{Boker97} found that the center of IC\,342 seems to be dominated by a young cluster of $\sim$10\,Myr. Although the optical and the NIR ranges trace stellar populations with different ages, these estimations seem to indicate that the star bursts in both sources have probably started at a similar time ago. 

Both NGC\,253 and IC\,342 show a very high density contrast with nearly two orders of magnitude between the Region I and the Region III. Furthermore, Regions III are rather small as compared with the Regions I, but are likely responsible for feeding the star formation at high rates in these galaxies. On the other hand, CO arises in this galaxies from gas with densities in the range $10^3-10^4\,cm^{-3}$ \citep{Israel03,Bayet04,Bayet06,Gusten06}. The average clouds in these two galaxies resemble the Sgr B2 star forming region in the GC. This cloud is centrally peaked surrounded by a large molecular envelope (\citealt{DeVicente96}) and shows that very recent star formation is still going-on in the very high density core (\citealt{DeVicente00}). As we have mentioned in Sect.~\ref{intro}, IC\,342  shows evolved star formation in the center, while at larger radii the stars seem to be younger \citep{Meier05}. We would like to emphasize that with single dish telescopes the overall scenario is a mixture of both cloud structures, where shock-dominated and PDR-dominated regions are not distinguished. Thus, the result is a mixture that leads to an intermediate stage of evolution \citep{Martin09b}, where still a large fraction of the gas is in a high density component and the cores are actively forming stars. As suggested by the detection of PDR tracers in NGC\,253 \citep{Martin09b} and illustrated in Fig.~\ref{nubes1}, photodissociation starts playing a role in these clouds in their outermost layers.
 
\subsection{Evolved starburst: M\,82}
The stellar clusters in the center of M\,82 are estimated to be the result of starbursts from at least $\sim{10-15}$\,Myr ago, as seen at optical wavelengths by \citet{Konstantopoulos09}. This value indicates that we are seeing a more evolved starburst (or post-starburst) in M\,82, older than the NGC\,253 and IC\,342 ones (we are not aware of a published stellar population age in the center of Maffei\,2). On the other hand, it is well established that newly formed stars in M\,82 create large photodissociation regions in its central region (\citealt{Martin06b}), which dominate over the other physical processes. Here, we follow the proposed overall scenario of M\,82 as a giant PDR, although we are developing for the future a more sophisticated interpretation where PDRs, XDRs and dense gas chemistry are mixed and will be better disentangle. This galaxy shows clouds with less density gradients than those found in the other galaxies, with a density contrast of only a factor of 6 between the two detected gas components (Regions II and III). This could be due to the fact that, in this evolved starburst galaxy, the H$_{\rm 2}$ density in Region II, of about $\sim$10$^5$ cm$^{-3}$, is surrounded by an envelope (not detected) where some molecules, especially the complex ones, are destroyed. The low density envelope surrounding the Region II could be dominated by mainly atomic composition with small column densities of resistant molecules like CO. In fact, \citet{Mao00} have detected CO in M\,82 with densities of $\sim$10$^{3.7}$ and $\sim$10$^3$ cm$^{-3}$, suggesting that this molecule does not trace the same material than high density tracers like CS. Likely the mechanical energy injected by the evolved starbursts had dispersed the outer parts of the clouds making larger and more diffuse  envelopes. Thus, the number of gas components with different densities and temperatures can be understood within the starburst evolutionary framework; in the later stage the dense molecular tracers are not arising from the outer parts, since they are dissociated in such regions. At this stage, the outermost photo-dissociated envelope, of densities $<10^5\rm cm^{-3}$, might represent a large fraction of the cloud, but with low abundance of complex molecules. At the same time, there is an important contribution of the Region III ($\sim$10$^6$ cm$^{-3}$), directly related with the large amount of star formation that have already been taken place. The high density component is also shown by observations of CH$_3$OH \citep{Martin06b}. 

Methyl acetylene (CH$_3$CCH) shows a very different behavior in M\,82 than the other molecules. First, its column density in M\,82, of $ 8.5\times10^{14}$\,cm$^{-2}$, is a factor of $\sim$4 larger than that of CS, and more than one order of magnitude than the HC$_3$N one. Its high fractional abundance with respect to $\rm H_2$, ${1.1\times10^{-8}}$, is puzzling for a molecule which is supposed to be easily dissociated in PDRs (\citealt{Fuente05}). The comparison with other molecular clouds dominated both by UV radiation and shocks, in the Galactic Center and in external galaxies, shows that M\,82 is the galaxy with the highest CH$_3$CCH abundance observed up to now. SgrA*\,(-30,-30) and G+0.18-0.04, used as templates of molecular clouds dominated by photodissociating radiation in the Galactic Center of the Milky Way (\citealt{Martin08a}), have relative abundances larger than a few 10$^{-9}$ (Mart\'in, 2006). On the other hand, the relative abundances we find in NGC\,253 and Maffei\,2 are $5.2\times10^{-9}$ and $2.0\times10^{-9}$ respectively. One possible explanation for the high CH$_3$CCH abundance in M\,82 is that this molecule can be created through gas phase ion-molecule or neutral-neutral reactions, while other molecules, like methanol or ethanol, are possibly only formed on ice-layers of dust grains (\citealt{Bisschop07}). When these complex molecules are dissociated by the strong UV fields of M\,82, their abundances decrease, while a similar decrease for CH$_3$CCH is balanced by ion molecule reactions favoring its formation. On the other hand, a strong anti correlation between $N_{\rm CH3CCH}$ and $T_{\rm kin}$ has been observed in galactic sources (\citealt{Lee96}), and the gas in M\,82 is indeed as cold as $\sim$30\,K. Methyl acetylene is not observed in the regions of densities $ <10^5\rm \,cm^{-3}$ since it would be dissociated, but in regions of intermediate densities (Region II), where it is well shielded from the dissociating radiation. Furthermore, we do not see it in the Region I, presumably because its abundance in such places is almost negligible, although it could exist there (\citealt{Churchwell83}).

\section{Conclusions}
\label{conclusions} 
From our multi-line studies of CS, HC$_3$N and CH$_3$CCH towards the centers of the starburst galaxies, M\,82, NGC\,253, IC\,342 and Maffei\,2, and using both LTE and non-LTE approaches, we  conclude that:

\begin{itemize}

\item CS and HC$_3$N are tracing gas components with similar physical properties. CS arises from a slightly denser gas (see Table~\ref{tab.LVG}), while HC$_3$N arises from warmer gas (see Table~\ref{tab.Boltmann}). Both molecules are excellent tools to derive the density structure of the average molecular clouds in galaxies. 

\item CH$_3$CCH arises from a ``warm" ($T_{\rm rot}$ from 13\,K up to 44\,K) gas components without large rotational temperature gradients. This molecule shows very high abundances and column densities in M\,82 ($1.1\times10^{-8}$\,cm$^{-2}$ and $ 8.5\times10^{14}$\,cm$^{-2}$ respectively), and probably arises from a region of densities $\sim$10$^5$\,cm$^{-3}$ which is well shielded from the strong UV radiation.
 
\item The gas traced by the observed molecules show differences between the galaxies in terms of density structure and composition. This fact allowed us to link the evolution of the starburst with the structure of the molecular gas clouds. First, Maffei\,2, the galaxy with the youngest starburst, shows a large low density halo (Region I $\sim$10$^4$\,cm$^{-3}$) and a more denser Region II with densities of $\sim$10$^5$\,cm$^{-3}$. This galaxy does not show signs of a dense gas core (Region III), probably because its starburst is still starting. Then, in an intermediate stage of evolution, NGC253 and IC342 clearly show three different gas components. First, a halo of low densities (Region I$\sim$10$^4$\,cm$^{-3}$), possibly surrounded by a small envelope where the molecules are dissociated. Then, an intermediate zone (Region II) of densities$\sim$10$^5$\,cm$^{-3}$; and finally a core of very dense gas (Region III$\sim$10$^6$\,cm$^{-3}$) which points out to the on-going star formation. Finally, M\,82, the more evolved starburst galaxy, shows a Region II of intermediate densities of $\sim$10$^5$\,cm$^{-3}$, probably surrounded by a large envelope dominated by photodissociating radiation, and a Region III of denser gas, comparable in size to the cores of NGC\,253 and IC\,342.

\end{itemize}

\begin{acknowledgements}
We thank the IRAM staff for their help with the observations and data reduction. R. Aladro acknowledges the hospitality of ESO Vitacura and the Joint ALMA Observatory. E.B. acknowledges financial support from the Leverhulme Trust.  We thank the referee for numerous and constructive comments. This work has been partially supported by the Spanish Ministerio de Ciencia e Innovaci\'on under project ESP2007-65812-C02-01, and by the “Comunidad de Madrid” Government under PRICIT project S-0505/ESP-0237 (ASTROCAM).
\end{acknowledgements}



\clearpage

\Online
\onecolumn
\begin{appendix}
\section{Table A.1. Gaussian fits parameters results.}
{
\begin{longtable}{c c l c c c c c c c l l}
\hline
Source	& ($\alpha$,$\delta$) & Line & $\int$ $T_{\rm{MB}}$\,dv	& FWHM		&$V_{\rm{LSR}}$	&$T_{\rm{MB}}$ & Rms & Int. Time & Ref.\\
        & ($'',''$) &    &  (K\,km\,s$^{-1}$)     & (km\,s$^{-1}$) & (km\,s$^{-1}$)   & (mK)      & (mK)         & (min)       \\[0.2cm]
\hline

NGC253\\
\hline
 & $(0,0)$& CS\,$(2-1)$ $^\bullet$	 &13.4$\pm$0.3	& 202.7	& 257.1	& 62.1	& 9.4 & 18.0 &  c\\
 & $(0,-6)$ & CS\,$(2-1)$ & 38.1$\pm$0.4 & 200.7 & 229.5 & 178.4 & 16.6    &  8.0 & a\\
 &$(0,0)$ & CS\,$(3-2)$ & 26.0$\pm$0.2	& 182.9	& 234.0	& 133.6	& 9.9 	& 100.0 &  c\\
 & $(0,0)$& CS\,$(4-3)$	& 24.2$\pm$0.4	& 178.4	& 236.9	& 127.7	& 13.5	& 119.0 &   c\\
 & $(0,-6)$ & CS\,$(5-4)$ & 18.6$\pm$0.7 & 135.6 & 180.2 & 129.5 & 11.3  & 16.0  & a\\
&  $(0,0)$& CS\,$(7-6)$ $^\dagger$ & 12.1$\pm$0.3	& 154.5	& 211.3	& 73.4	& 8.2 & 90 &   b\\

\cline{2-12}
 & $(0,-9)$& HC$_3$N\,$(9-8)$  & 5.8$\pm$0.6  & 63.0 $^\star$ & 184.0 $^\star$ & 86.0 & -  & - &  d \\
  & $(0,-9)$& HC$_3$N\,$(10-9)$  & 5.3$\pm$0.3 & 63.0 $^\star$ & 184.0 $^\star$ & 80.0 & -  & - &  d\\
 & $(0,-9)$& HC$_3$N\,$(12-11)$  & 4.4$\pm$0.7 & 63.0 $^\star$ & 184.0 $^\star$ & 66.0 & -  & - & d \\
 & $(0,-9)$& HC$_3$N\,$(15-14)$  & 3.6$\pm$0.6 & 63.0 $^\star$ & 184.0 $^\star$ & 54.0 & - & - &  d \\
 & $(0,-9)$& HC$_3$N\,$(17-16)$  & 3.4$\pm$0.5 & 63.0 $^\star$ & 184.0 $^\star$& 50.0 & -  & - & d \\
 & $(0,-9)$& HC$_3$N\,$(26-25)$  & 3.2$\pm$0.7 & 63.0 $^\star$ & 184.0 $^\star$& 47.0 & -  & - &  d \\
\cline{2-12}
 & $(0,0)$&  CH$_3$CCH\,$(8_0-7_0)$  & 3.4$\pm$0.1 & - & - & - & -  & - & c \\
 & $(0,0)$& CH$_3$CCH\,$(9_0-8_0)$  & 4.3$\pm$0.2 & - & - & - & -  & - & c \\
 & $(0,0)$&  CH$_3$CCH\,$(10_0-9_0)$  & 5.1$\pm$0.5 & - & - & - & -  & - &  c\\
 & $(0,0)$& CH$_3$CCH\,$(13_0-12_0)$ $^\dagger$   & 2.9$\pm$0.3 & 155.0 &  250.0 & 17.3 & 3.8 & 90.0 &  a\\[0.2cm]
\hline
M82\\
\hline
 & $(+13,+7.5)$& CS\,$(2-1)$	&9.4$\pm$0.2	& 102.6	 & 300.1 & 86.3	& 4.6 	& - &  b\\
 &$(+13,+7.5)$ & CS\,$(3-2)$ 	& 11.5$\pm$0.2	& 105.6	& 293.8	& 103.7	&  2.6	& 481 & a\\
 & $(+13,+7.5)$& CS\,$(4-3)$	& 6.5$\pm$0.8	& 87.7	& 309.2	& 70.1	& 16.6 	& - &  b\\
 & $(+13,+7.5)$& CS\,$(5-4)$	 & 4.2$\pm$0.2	& 71.6	& 316.4	& 55.5	&  7.3	& 248  & a\\
&  $(+13,+7.5)$& CS\,$(7-6)$ $^\dagger$ & 0.9$\pm$0.1	& 53.0	& 317.8	& 16.7	& -	& -  & b\\
\cline{2-12}
 & $(+13,+7.5)$   & HC$_3$N\,$(15-14)$ &  1.2$\pm$0.2	& 97.3	& 308.8	& 11.5   & 2.5    &  59   &   a\\
 & $(+13,+7.5)$   & HC$_3$N\,$(16-15)$ &  1.0$\pm$0.1	& 100.0 $^\star$ & 300.0 $^\star$ &  9.6   & 1.4  &  292 &   a\\
 & $(+13,+7.5)$  & HC$_3$N\,$(17-16)$ &  1.3$\pm$0.1	& 96.9	& 314.3	&   13.0   & 2.4  &   158 &    a\\
 & $(+13,+7.5)$   & HC$_3$N\,$(18-17)$ &  0.8$\pm$0.1	& 87.5 & 308.5 &  8.6	   & 2.0  &  217 &    a\\
& $(+13,+7.5)$   & HC$_3$N\,$(24-23)$ &  0.4$\pm$0.3	& 42.6	& 333.8 &  9.9	   & 4.6 &  294 &   a\\
& $(+13,+7.5)$   & HC$_3$N\,$(28-27)$ &  $\leq$0.2		& 100.0 $^\star$ & - & -  & 1.1 &  1400 &    a\\
\cline{2-12}
& $(+13,+7.5)$   & CH$_3$CCH\,$(6_0-5_0)$  &  3.7$\pm$0.1	& 123.8	& 315.0 &  28.1   & 1.0     &   168 &    h\\
& $(+13,+7.5)$   & CH$_3$CCH\,$(8_0-7_0)$  &  7.4$\pm$0.6	& 122.9	& 310.0 $^\star$&  56.8   & 2.5     &   59 &    a\\
& $(+13,+7.5)$   & CH$_3$CCH\,$(9_0-8_0)$  &  7.1$\pm$0.1	& 109.6 & 322.6	&  63.9   & 1.7   &  158  &  a\\
& $(+13,+7.5)$   & CH$_3$CCH\,$(10_0-9_0)$  &  8.7$\pm$0.2	& 97.5 & 325.0	&  84.1   & 3.0   &  281  &  a\\
& $(+13,+7.5)$   &  CH$_3$CCH\,$ (15_0-14_0)$  &  7.4$\pm$0.7& 85.1	& 339.1	&  81.2   & 8.0 & 56 &   a\\
& $(+13,+7.5)$   &  CH$_3$CCH\,$(16_0-15_0)$  &  2.4$\pm$0.1& 70.1	& 337.5	& 32.1    &  3.1  &  799    & a\\
\hline
IC342\\
\hline
 & $(0,0)$& CS\,$(2-1)$ & 5.0$\pm$0.1  & 53.9 & 31.7 & 87.4	& 2.5  & 28 &  a\\
& $(0,0)$&  CS\,$(3-2)$ & 4.9$\pm$0.1 & 51.7 & 31.2 & 88.5 & 3.0  & 28 &  a \\
& $(0,0)$&  CS\,$(5-4)$ & 3.4$\pm$0.1  & 63.4 & 38.2 & 37.6 & 2.9  & 57 & a \\
& $(0,0)$&  CS\,$(7-6)$ $^\dagger$ & 1.0$\pm$0.4  & 75.6 & 29.3 & 12.3 & 8.5 & 50 & b\\

\cline{2-12}
& $(0,0)$&  HC$_3$N\,$(9-8)$ & 0.7$\pm$0.1   & 54.4 & 34.4 & 12.2 & 1.2  &  126 & a \\
& $(0,0)$&  HC$_3$N\,$(10-9)$ & 0.7$\pm$0.1  & 40.9    & 35.5     & 15.3 & 2.8 &   84 & a \\
& $(0,0)$&  HC$_3$N\,$(11-10)$ & 0.7$\pm$0.1 & 58.7    & 34.0     & 12.0 & 1.0 &   218&  a \\
& $(0,0)$&  HC$_3$N\,$(12-11)$ & 0.7$\pm$0.1 & 52.7    & 39.3     & 12.4 & 1.3 &   77 &  a \\
& $(0,0)$&  HC$_3$N\,$(15-14)$ & 0.5$\pm$0.1 & 50.9    & 39.2     & 9.8  & 2.0 &   56 & a \\
& $(0,0)$&  HC$_3$N\,$(16-15)$ & 0.4$\pm$0.1 & 50.0 $^\star$ & 31.0 $^\star$ & 6.7 & 1.8 &  190 &  a \\
& $(0,0)$&  HC$_3$N\,$(17-16)$ & 0.7$\pm$0.1 & 101.3 & 44.1 & 6.6 & 1.5 &  416 & a \\
& $(0,0)$&  HC$_3$N\,$(24-23)$ & 0.7$\pm$0.2 & 76.7   & 12.7     & 8.4  & 2.8 &  211 & a \\
\cline{2-12}
& $(0,0)$&   CH$_3$CCH\,$(5_0-4_0)$ & 0.2$\pm$0.1 & 36.8      & 33.0 &3.9 & 5.8  & 278 & a \\
& $(0,0)$&   CH$_3$CCH\,$(6_0-5_0)$ & 0.3$\pm$0.1 & 60.0 $^\star$ & 38.6 & 4.5& 6.7  & 276 & a \\
& $(0,0)$&  CH$_3$CCH\,$(8_0-7_0)$ & 0.7$\pm$0.1 & 50.0 $^\star$ & 25.5 &13.7 & 1.5  & 56 & a \\
\hline
Maffei2\\
\hline
 & $(0,-7)$& CS\,$(2-1)$ $^@$ & 0.9$\pm$0.1  & 132.9 & -19.0 & -& - & - & f \\
& $(0,-4)$&  CS\,$(3-2)$ & 6.1$\pm$0.5 & 175.9  & -36.0 & 33 & - & - & g  \\
 & $(0,0)$& CS\,$(5-4)$ & 3.0$\pm$0.3  & 139.2 & -50.0 $^\star$ & 20.3 & 3.1 & 99 &  a \\

\cline{2-12}
& (0,0)&  HC$_3$N\,$(9-8)$ & 1.0$\pm$0.1   & 130.0 $^\star$ & -33.3 & 7.6 & 78.1  &  112 &  a \\
& $(0,0)$&  HC$_3$N\,$(12-11)$ & 1.8$\pm$0.4   & 112.2 & -44.6 & 15.3 & 3.9  &  21 &  a\\
& $(0,0)$&  HC$_3$N\,$(15-14)$ & 0.4$\pm$0.1   & 130.0 $^\star$  & -50.0 $^\star$  & 3.1 & 1.0  &  112 & a \\
& $(0,0)$&  HC$_3$N\,$(23-22)$ & $\le$0.3   & 130.0$^\star$ & - & - & 2.2  & 183 & a \\
& $(0,0)$&  HC$_3$N\,$(28-27)$ & $\le$0.2   & 130.0$^\star$ & - & - &  8.5 & 140 & a \\

\cline{2-12}
& $(0,0)$&  CH$_3$CCH\,$(5_0-4_0)$ & 0.3$\pm$0.1  & 142.5 & -50.0 $^\star$ & 2.1 & 1.0  & 224  & a \\
& $(0,0)$&  CH$_3$CCH\,$(6_0-5_0)$ & 0.8$\pm$0.1  & 146.3 & -50.0 $^\star$ & 4.8  & 1.0  & 224  &  a \\
& $(0,0)$&  CH$_3$CCH\,$(8_0-7_0)$ & 0.7$\pm$0.1   & 130.0 $^\star$  & -60.4 & 5.3 & 1.1  & 112  & a \\
& $(0,0)$&  CH$_3$CCH\,$(13_0-12_0)$ & $\le$0.3   &130.0 $^\star$  & - & - & 3.0  & 224 & a \\
& $(0,0)$&  CH$_3$CCH\,$(14_0-13_0)$ & $\le$0.2   & 130.0 $^\star$ & - & - & 2.0  & 253 & a \\

\hline

\caption{All lines were observed with IRAM 30-m telescope except: $^\bullet$: line observed with SEST; $\dagger$, line observed with JCMT; $^@$, line observed with 12-m NRAO. A $^\star$ following the FWHM or (/and) $V_{\rm{LSR}}$ means that the parameter was fixed during the Gaussian fit. Lines with values of only rms and integration time are not detections, but upper limits. References: (a): this work; (b): \citet{Bayet09}; (c): \citet{Martin06b}; (d): \citet{Mauersberger90}; (e): \citet{Mauersberger91}; (f): \citet{Sage90}; (g): \citet{Mauers89b}; (h):Aladro et al, 2009b, in prep.}
\label{TableA1}
\end{longtable}
}

\end{appendix}

\end{document}